\documentclass[preprint,12pt]{elsarticle}

\usepackage{graphicx}
\usepackage{xcolor}
\usepackage{float}

\usepackage{array}  
\usepackage{amssymb}
\usepackage{amsthm}
 \usepackage{amsfonts,amssymb}
\usepackage{amsmath}

\def\a{\alpha}
\def\b{\beta}
\def\ep{\epsilon}
\def\eps{\epsilon}

\def\g{\gamma}

\def\g{\gamma}

\theoremstyle{definition}
\newtheorem{definition}{Definition}

\journal{Annals of Physics}

\begin{document}

\begin{frontmatter}


\title{On the pseudo-Riemann's quartics in Finsler's geometry}



\author{Yakov Itin}

\address{Mathematics Department, Jerusalem College of Technology; Jerusalem, Israel}

\begin{abstract}
An extension of Riemmann's geometry into a direction dependent geometric structure is usually described by Finsler's geometry. Historically, this construction was motivated by the well-known  Riemann's quartic length element example. Quite surprisingly, the same  quartic expression emerges in solid-state electrodynamics as a basic dispersion relation---covariant Fresnel equation. Consequently,  Riemann's quartic length expression can be interpreted as  a mathematical model of a well-established physics phenomena. In this paper, we present various examples of Riemann's quartic that demonstrate that  Finsler's geometry is too restrictive even  in the case of a positive definite Euclidean signature space. In the case of the spaces endowed with an  indefinite (Minkowski) signature, there are much more singular hypersurfaces where the strong axioms of Finsler's geometry are  broken down. We propose a weaker definition of Finsler's structure that is required to be satisfied only on  open subsets of the tangent bundle. We exhibit the characteristic  singular  hypersurfaces related to Riemann's quartic and briefly discuss  their  possible physical interpretation.

\end{abstract}


\end{frontmatter}


\section{Introduction}
In 1854,  Riemann  developed his theory of higher dimension geometry and delivered in 1854 his famous Habilitation lecture at G\"{o}ttingen  entitled {\it "Über die Hypothesen welche der Geometrie zu Grunde liegen"} (On the hypotheses which underlie geometry).   His professor and examiner,   Gauss, was  greatly impressed by the ideas presented in this work. 
The paper was only published  by Dedekind in 1868, two years after Riemann's  death. 

Riemann wrote: {\it "$ds$ is the square root of an always positive integral homogeneous function of the second order of the quantities $dx$."} In other words, Riemann's construction is started with the quadratic length element 
\begin{equation}\label{0-FRie}
ds = \sqrt{g_{ij}(x) dx^{i} dx^{j}}\,,
\end{equation}
where $g_{ij}(x)$ is a symmetric positive definite  tensor. 
 Nowadays this length element   serves as a well-known  basis of Riemannian geometry. Its indefinite version  with the tensor $g_{ij}(x)$ of the Lorentz signature  provides a mathematical background of General Relativity.  
 
In his Habilitation lecture, Riemann  continued: 
{\it "...The
next case in simplicity includes those manifolds in which the line-element may
be expressed as the fourth root of a quartic differential expression. The investigation of this more general kind would require no really different principles, but would
take considerable time and throw little new light on the theory of space, especially as
the results cannot be geometrically expressed, I restrict myself, therefore, to those
manifolds in which the line-element is expressed as the square root of a quadratic differential expression..."} 
We denote the quartic Riemann's expression as 
\begin{equation}\label{1-FRie}
ds=\sqrt[4]{M_{ijkl}(x) dx^{i}dx^{j}dx^{k}dx^{l}}\,.
\end{equation}

The well-known extension of the  quartic Riemann's line element is Finler's geometry that deals with a general  expression for the line elements   
\begin{equation}\label{2-FRie}
    ds=F(x, dx).
\end{equation}
Here $F(x, dx)$ is an arbitrary smooth positive function that is  first-order homogeneous in its second argument. 
After intensive study of Finsler, Cartan, Chern, and many others, this subject turned into a solid mathematical theory, see \cite {Rund(1959)}, \cite{Chern(2005)} and the references given therein.  The classical Finsler's geometry is restricted to the positive defined case, i.e., to the Euclidean signature. 

Various physics applications motivate study of Finsler geometry  in the {\it pseudo-Euclidean} domain.   The well-known example  from classical physics is the Randers  metric that can be viewed as an attempt of  geometrization of the electric charges motion.
 Possible pseudo-Finsler modifications  of gravity were  discussed in \cite{Lammerzahl:2018lhw}. Also in quantum field theories, Finsler's geometry is a rather popular construction. In particular, in  \cite{Kostelecky:2011qz}, \cite{Edwards}  pseudo-Finsler geometry is applied as a Lorentz symmetry violating model. In \cite{Itin:2014uia} Finsler-type modification of the Coulomb law and corresponding additional splitting of  hydrogen energy levels was calculated. This result provides a way for an  experimental verification of Finsler's structure. 
Another example is the covariant premetric dispersion relation in electromagnetism \cite{Hehl-Obukhov(2003)} that is presented as a general fourth order polynomial  equation. Due to the well-observed birefringence effect in anisotropic optics, the covariant premetric electromagnetism must be considered as a viable theory. Remarkably, the covariant premetric dispersion relation has exactly the same form as the one exemplified by Riemann in Eq.(\ref{1-FRie}). 
 
 This paper is aimed to discuss  how Riemann's quadric
 \begin{equation}\label{Riem-quadr}
     Q(x,v)=M_{ijkl}(x) v^{i}v^jv^kv^l
 \end{equation}
 can be squeezed into the basic  definitions of Finsler's metric. Also we are interested in the question: Which additional information of the wave propagation can be derived  from the knowledge of Finsler's structure. 
 Notice that in order to deal with indefinite  signatures, we cannot  restrict  the expression (\ref{Riem-quadr}) to be positive-definite (Euclidean). 
 
 The paper is organized as follows: In Sect. 2, we provide a brief account of premetric dispersion relation in electromagnetism. The covariant version of the tensor $M_{ijkl}$   is shown to be expressed via the observable physical parameters. In Sect. 3, we provide the classical definition of Finsler's geometry and discuss the reasoning for its strong restriction.  
 In Sect. 4, we apply Finsler geometry in the Euclidean domain to Riemann's quadric. We observe that even in this classical subject, the Finsler metric is degenerated on three-dimensional surfaces of tangential vector space. So the basic definition must be modified.  In Sect. 5, we discuss the extension of Finsler geometry into the non-Euclidean domain. In contrast to the familiar pure Lorentzian case, this  construction is well-defined  and differentiable only in a certain {\it conical subset of the tangent space.} 
 Sect. 6 is devoted to a novel weakened definition of Finsler structure. We require this structure to be defined on an open subspace of the tangent bundle. We  explicitly identify the singular subsets where some of the restrictions of the basic definition are broken down. In our opinion, these subsets provide important information about the physics of the corresponding processes. The case of an uniaxial crystal is aimed to give a physical meaning to the formal structure definition.  
In the Conclusions section, we discuss the possible extensions and applications of the weekend Finsler's structure.  The Appendix section is devoted to various examples of  explicit calculations of Finsler's metric in the two-dimensional  case.
 
 In this paper, we are dealing with 4-dimensional manifolds $M$, i.e.,  a connected paracompact
differentiable manifold of dimension four and of the smoothness class $C^\infty$. Roman indices are changed over the range $i,j, \cdots=0,1,2,3$. The 
Einstein summation convention is applied throughout:  summation for two  indexes appearing  in  upper and lower positions is implicitly implied.
Being interested here in the directional dependence of Finsler's structure we suppress in most cases the functional dependence of the variable $x$ and denote the tangential space $T_xM$ at a general unspecified point $x\in M$ by $V$. 

\section{Riemann's quartic in electromagnetism}
The quartic construction was considered by Riemann  as only a formal possibility. It is a quite remarkable fact that this construction  can be given a well-posed physical interpretation in electromagnetic wave propagation physics. As it was already known to Gauss and Riemann,  the geometry of our world can be probed by the light propagation phenomena, i.e., by the electromagnetic waves. Consequently,  the geometric structure  can be viewed as only a secondary phenomena that is established by the properties of the media. For  details of this {\it premetric approach},  see \cite{Hehl-Obukhov(2003)} and \cite{Itin(2009)}.

Consider the source-free Maxwell system 
\begin{equation}\label{FEq}
    dF=0\,,\qquad dH=0
\end{equation}
Here $F$ is the {\it field strength} 2-form, while $H$ the {\it excitation} tensor density 2-form, $d$ states for the exterior derivative operator. 
In coordinates,  two basic differential forms are expressed as 
\begin{equation}
    F=\frac 12 F_{ij}dx^i\wedge dx^j\,, \qquad 
H=\frac 12 H^{ij}\epsilon_{ijkl}dx^k\wedge dx^l\,.
\end{equation}
Now $F_{ij}$ is termed as the field strength tensor, and $H^{ij}$ as the excitation pseudo-tensor. The quantity   $\epsilon_{ijkl}$ is the Levi-Civita  permutation pseudotensor. 

In a wide range of the field's magnitudes,  a linear {\it constitutive relation} between  two second-order tensors, $F$ and $H$, can be assumed. In coordinates components, this relation is written as 
\begin{equation}\label{const}
    H^{ij}=\frac 12 \chi^{ijkl}F_{kl}\,.
\end{equation}
 Here,  $\chi^{ijkl}$ is a {\it constitutive pseudo-tensor} that describes the  physical parameters of media. The field equations (\ref{FEq}) together with the constitutive relation (\ref{const}) form  a well-posed system of partial differential equations. This system is  completely independent of the geometric structure of the manifold. In particular, it does not involve the metric tensor. 
Moreover, the system of electromagnetic field equations itself can be interpreted  as a manifestation of some special interior  geometry  encoded  in the constitutive tensor  $\chi^{ijkl}$. 
In an isotropic media, as well as in vacuum,  the  constitutive tensor is presented by the tensor square of a  second order tensor $g^{ij}$
\begin{equation}\label{Lor}
    \chi^{ijkl}=\frac 12\sqrt{-g}\left(g^{ik}g^{jl}-g^{il}g^{jk}\right)\,.
\end{equation}
In this case, the characteristic equation for existence of the  electromagnetic waves (the dispersion relation) is given in the reduced quartic form  
\begin{equation}\label{Lor1}
    (g^{ij}q_iq_j)^2=0\,,
\end{equation}
where $q_i$ is the four-dimensional wave covector. 
With the Minkowski metric tensor $g^{ij}={\rm diag} (1,-1,-1,-1)$, the left-hand-side of Eq.(\ref{Lor1})  gives Lorentz square norm of the wave covector $q_i$.  This case can be interpreted as  a fact that a specific constitutive tensor of vacuum electromagnetism (\ref{Lor}) yields the  pseudo-Riemannian geometry on  space-time. The standard Euclidean geometry emerges on  three-dimensional subspaces of the space-time manifold. 

In solid state physics,   a general constitutive tensor $\chi^{ijkl}$ includes the full set of 36 independent components. In $(1+3)$-splitting form, it is represented as \cite{Hehl-Obukhov(2003)}
   \begin{equation}\label{gen-8}
 \chi^{ijkl}=\left( \begin{array}{cc}
\ep^{\a\b} & \g^\a{}_\b \\
\tilde{\g}^\a{}_\b&\pi_{\a\b} 
 \end{array} \right)\,.
\end{equation}
Here the four-dimensional tensor $\chi^{ijkl}$ is described by four three-dimensional block matrices numerated by the Greek indices changing in the range $\a,\b=1,2,3$. 
The block matrices $\ep^{\a\b} $  and $\pi_{\a\b}$  describe the pure physical parameters called  {\it permittivity} and the {\it impermeability} of the media,respectively.   
 The mixed-type tensors $\g^\a{}_\b$ and $\tilde{\g}^\a{}_\b$  are the {\it electric-magnetic cross-terms}. In other words, the components of  the constitutive tensor $\chi^{ijkl}$ have a well-posed physical interpretation. At least part of them are observed in solid-state physics experiments.

 Similarly to the isotropic case (\ref{Lor1}), the electromagnetic wave propagation in media is described by a wave covector $q_i$ that  satisfies some special algebraic   conditions  (dispersion relations).  For a compact expression of these conditions, it is useful to define the  left-right-dual of the tensor $\chi^{mnkl}$ 
 \begin{equation}
     \tilde{\chi}_{acbd}=\frac 14 \eps_{abmn}\chi^{mnkl}\eps_{klcd}
 \end{equation}
 and the fully symmetric tensor $G^{ijkl}=G^{(ijkl)}$
\begin{equation}
    {G}^{ijkl}[\chi]=\frac{1}{3!}\, \chi^{a(i j|b}\,\tilde{\chi}_{acbd}\,\chi^{c|kl)d}\,.
\end{equation}
The dispersion relation is  given now by a  quartic equation, \cite{Hehl-Obukhov(2003)}, \cite{Itin(2009)}
\begin{equation}\label{gen-9}
G^{ijkl}q_iq_jq_kq_l=0\,,
\end{equation}
that is termed as  {\it covariant  Fresnel equation}. 
 Similarly to the isotropic case, the natural problem appearing here is: 

\noindent {\it Which geometric structure can be related to the fourth-order polynomial expression $G^{ijkl}q_iq_jq_kq_l$?}

Most remarkable, that the covariant  Fresnel equation is exactly of the same form as appearing in the quartic Riemann's length expression. The only difference is that in Riemann's construction, the quartic is assumed to be positive defined. For electromagnetic quartic (\ref{gen-9}), this restriction must be lifted.

\section{Finsler's construction}
A  generalization of Riemann's  quadratic (\ref{0-FRie}) and quartic  (\ref{2-FRie}) linear elements   to a  general linear element was evaluated by Finsler (1918). In this section, we briefly discuss the basic definitions of Finsler's geometry. For the details, see, e.g.,  \cite{Rund(1959)}, \cite{Chern(2005)} and the references given therein.
\subsection{Basic definitions}
 
\begin{definition}\label{def1}
{\it Finsler manifold} is a differentiable manifold $M$ endowed with a {\it Finsler function} $F(x,v)$ of two different sets of variables: a point $x\in M$ and a vector $v\in T_xM$. Finsler's function is assumed to satisfy the following conditions: 
\begin{itemize}
\item[(F1)] {\it Continuity:} 
$F(x,v)$ is defined and continuous on the  entire tangent bundle $TM$, i.e., for all $x$ and $v$.
\item[(F2)] {\it Positive definiteness:} For all $x$ and all $v$, Finsler's function is non-negative,  $F(x,v) \ge 0$.  Equality holds only at the origin  $v = 0$.
\item[(F3)] {\it Positive homogeneity:} For all $\lambda > 0$ and all $v \ne 0$, the first order homogeneity equation holds,
\begin{equation}
    F(x,\lambda v) = \lambda F(x,v).
\end{equation}
\item[(F4)] {\it Smoothness:} For all points $x$ and for all {\it nonzero} vectors $v$, the function $F(x,v)$ is smooth in its  second argument (at least to the class $C^4$). 
\item[(F5)]  {\it Finsler's metric:} 
    Finsler's metric is defined as the Hessian of the Finsler function square $F(x,v)^2$
\begin{equation}\label{Hess0}
f_{ij}(x,v)=
\frac 12\,\frac{\partial^2  F(x,v)^2 }{\partial v^{i}\partial v^{j}}\,.
\end{equation}
The metric is required to be non-degenerate for  all points $x$ and  for all  vectors $v\ne 0$.
\item[(F6)]  {\it Euclidean signature:} 
The signature of the metric $f_{ij}(x,v)$ is assumed to be Euclidean. It means that at every point $x\in M$ there is a basis where the metric is equal to the unit matrix   $f_{ij}(x,v)={\rm diag}(1,1,1,1)$. 

\end{itemize}
\end{definition}

\subsection{Motivations and comments} 
Some motivations for the  conditions presented in the definition of the Finsler structure  can be given as follows. The first group of conditions describes the norm (length) of an arbitrary vector $v\in T_xM$  at an arbitrary point $x\in M$. 

\vspace{0.2cm}

(F1) Continuity condition  guarantees that a   line element (length of a vector) $F(x,v)$ is defined (and final) for every vector $v$ attached  at an arbitrary point $x$. For the neighborhood   vectors, Finsler's function  provides close lengths. 

(F2)  Positiveness of $F(x,v)$ guarantees a positive length for an arbitrary non-zero vector $v$. 
 In particular, for a smooth curve $x(s)\in M$ with a  tangent vector $v=\dot{x}(s)$, Finsler function $F(x, v)$  assigns  a positive length
\begin{equation}\label{eq:length}
\ell = \int _{s_1}^{s_2} F \big( x (s), \dot{x}(s) \big) \, ds\,.
\end{equation}

(F3) The positive  homogeneity of the degree one provides proper physical dimension of the length function. Moreover, it guarantees that the length of an arbitrary curve is   invariant  under  positive  oriented reparametrization of the curve. In order to permit asymmetric structures, this condition  is required only for the  positive values of the parameter $\lambda$. In other words,  $F(x, -v)\ne F(x,v)$ in general. 

\vspace{0.2cm}
The second group of conditions provides an additional structure on the tangential vector space $T_xM$ (and on the total manifold $M$). It introduces a {\it scalar product of two vectors depending on a specially chosen vector} $v$. The corresponding metric tensor (even being depended on $v$) opens a room for applying the standard tools of differential geometry---connections, curvature tensors, and so on. 
\vspace{0.2cm}

(F4) The smoothness condition permits the derivatives of the function $F(x,v)$ to be singular only for $v=0$. This restriction is necessary because already in the simplest case (\ref{0-FRie}), the derivative of the square root expression is not defined for $v=0$. One formulates the condition as differentiability of $F(x,v)$ on a subset of a bundle  $TM\backslash \{0\}$, where $\{0\}$ is a null section.  In fact, only the derivatives of the Lagrangian  $L(x,v)=F(x,v)^2$ are in use here. Thus it is possible to require a bit weaker condition---smoothness of the square $F(x,v)^2$ instead of the function $F(x,v)$ itself.    Such modification is useful for Riemann's quadratic expression.     We will see that  this condition is related to the Euclidean (elliptic) structure on the manifold. In the pseudo-Euclidean case, it must be modified to permit additional singular sets. 

(F5) 
The Euclidean metric construction can be written in a coordinate-free form:  \\
For each nonzero $v\in T_pM$, the Hessian of $F^2(x,v)$ at some $v$ is applied at two independent vectors $u$ and $w$,
\begin{equation}\label{Hess}
    {\displaystyle \mathbf {f} _{v}(u,w):={\frac {1}{2}}\left.{\frac {\partial ^{2}}{\partial s\partial t}}
F(x,v+su+tw)^{2}\right|_{s=t=0}}
\end{equation} 
As a result, the metric function turns out to be a homogeneous function of the zero order. 
\begin{equation}
    f_{ij}(x,\lambda v)=f_{ij}(x, v).
\end{equation}
(F6) 
The Euclidean signature condition require the matrix $f_{ij}$ to be positive definite for an arbitrary  vector $u$
\begin{equation}
   f_{ij}(x,v)u^iu^j\ge 0. 
\end{equation}

Under a general linear transformation of the basis, the components of the matrix  $f_{ij}$ are transformed by the standard tensor rule. Thus $f_{ij}$ is referred to as {\it Finsler's  metric.} This metric-type tensor depends not only on the  point $x\in M $ but also on the directions given by the tangent vector $v\in T_xM$.  Recall that $f_{ij}(x,v)$ is not necessarily defined for $v=0$. 
Using Finsler's metric, the scalar product of two arbitrary vectors $u$ and $w$ can be defined in the standard way  
\begin{equation}
    (u,w)|_{f(v)}=f_{ij}(x,v)u^iw^j\,.
\end{equation}
For a given Finsler's function, this expression can be treated as a {\it scalar product of vectrors} $u,w$  {\it with respect to the chosen vector} $v$. 
Correspondingly, a {\it norm of a vector $u$ with respect to the chosen vector} $v$ is defined as 
\begin{equation}\label{norm}
   |u|_{f(v)}^2= (u,u)|_{f(v)}=f_{ij}(x,v)u^iu^j\,.
\end{equation}
Due to the homogeneity condition, the norm of the vector $v$ calculated with respect to the  vector $v$ itself is equal to the square of the Finsler function
\begin{equation}
    |v|_{f(v)}^2= (v,v)|_{f(v)}=f_{ij}(x,v)v^iv^j=F(x,v)^2\,.
\end{equation}
Deviation of the Finsler structure from the Riemann one is characterized by the {\it Cartan tensor} 
\begin{equation}
    C_{ijk}=\frac 12\frac{\partial} {\partial v^{k}} f_{ij}(x,v)= \frac 14 \frac{\partial^3 F(x,v)^2}{\partial v^{i}\partial v^{j}\partial v^{k}}\,.
\end{equation}
When $C_{ijk}=0$ for all vectors $v$ and for all points $x$, the metric $g_{ij}$ is independent of the direction argument, hence the space is pure  Riemannian. 

Note that most of Finsler's functions, in particular those  that are derived from pseudo-Riemann quartics, provide  metrics that are degenerate on some subset of the vectors $v$. 
In mathematical texts, the metric condition is  extended to a weaker   subadditivity inequality  condition. It is a  generalization of the standard triangle inequality of Euclidean geometry.

    (${\rm F5}^\backprime$--${\rm F6}^\backprime$) {\it Subadditivity:} 
    \begin{equation}
        F(x,v + w)\le F(x,v) + F(x,w)\quad  \mbox {for all} \quad x\in M \quad \mbox {and all } \quad v,w\in T_xM .
    \end{equation}
In this case, one does not need the metric construction at all. This extension, however, does  not conform  with the standard construction of Riemann's geometry (metric, connection, curvature, etc.). Moreover, it  cannot be extended to the pseudo-Riemannian case. 

\section{Finsler's function derived from Riemann's quartic}
Since we are dealing with Riemann's quartic as a basic building element, our first task is to construct a proper  first order homogeneous Finsler's  function $F(x,v)$ related to the fourth order homogeneous function $Q(x,v)$. The questions are:
\begin{itemize}
    \item[(i)] Is it  always possible  to construct a  function $F(x,v)$ from an arbitrary general function $Q(x,v)$?
    \item[(ii)] Is such function $F(x,v)$  unique, or there is a family of possible functions $F(x,v)$ related to the same quartic $Q(x,v)$?
    \item[(iii)] Is it  possible to choose the best function $F(x,v)$ representing $Q(x,v)$?
\end{itemize}
We start with a simpliest case of positive definite quartic. 

{\it Positive definite quartic:} Let  the  quartic satisfy the inequality  
\begin{equation}\label{EuclQuartd0}
  Q(x,v)> 0,  
\end{equation}
for all nonzero $v\in T_xM$ and for all $x\in M$. Additionally, $Q(x,0)=0$. Consequently, Finsler's function can be constructed in a unique form 
\begin{equation}\label{EuclQuartd1}
  F(x,v)=\sqrt[4]{Q(x,v)}.
\end{equation}
For  the general Riemann quartic 
\begin{equation}\label{EuclQuartd0}
 Q(x,v)=M_{ijkl}(x) v^{i}v^{j}v^{k}v^{l}. 
\end{equation}
   $M_{ijkl}$ is restricted to a positive  definite tensor such that the inequality  
   \begin{equation}\label{EuclQuartd0}
 M_{ijkl}(x) v^{i}v^{j}v^{k}v^{l}> 0,  
\end{equation}
 holds  for all nonzero vectors  $v\in T_xM$. In this case,  Finsler's function is defined uniquely as 
\begin{equation}
    F(x,v)=\sqrt[4]{M_{ijkl}(x) v^{i}v^{j}v^{k}v^{l}}\,.
\end{equation}
This function evidently satisfying the conditions (F1)--(F4). We will discuss the additional metric  conditions  (F5)--(F6) later. 

{\it Indefinite quartic:}
We consider an indefinite function $Q(x,v)$ that changes it sign on the tangent space $T_xM$ and vanishes on some subset of this vector space. There are three possibilities for the function $F(x,v)$.

(1) The function 
\begin{equation}\label{sect5-2}
    F(x,v)=\sqrt[4]{Q(x,v)}=\sqrt[4]{M_{ijkl}v^iv^jv^kv^l}
\end{equation}
is defined and continuous only for a set of vectors $v$ that satisfy the relation $Q(x,v)>0$. In physics literature 
\cite{Asanov(1985)}, it was proposed to deal with such a restricted part of the tangent space  taking into account only the set of ``admissable''  timelike vectors. As it is well-known, the curves with timelike tangents  describe in General Relativity the motion of massive particles.   In this construction, it is possible to preserve almost all basic features of the Finsler structure but only in a restricted conic region. 

(2) Following \cite{Beem(1970)}  we can define  the absolute value form 
\begin{equation}\label{sect5-3}
    F(x,v)=\sqrt[4]{|Q(x,v)|}=\sqrt[4]{|M_{ijkl}v^iv^jv^kv^l|}
    \end{equation}
   that allows us to deal with the entire tangent space. It has, however, a clear disadvantage in giving the same positive values of Finsler's function for positive and negative values of the quartics $Q(x,v)$. In other words, this Finsler's function forgets the initial sign of the quartic $Q(x,v)$. 
    
(3) We define the Lagrangian as follows
\begin{equation}\label{sect5-4}
    F(x,v)=\sqrt[4]{|Q(x,v)|}\,{\rm sgn} Q(x,v)=
      \begin{cases} 
      \,\,\,\,\sqrt[4]{|Q(x,v)|} & {\rm for}\,\,  v^i \,\, {\rm satisfying}\,\,   Q(x,v)\ge 0\\
      -\sqrt[4]{|Q(x,v)|} & {\rm for}\,\,  v^i \,\, {\rm satisfying}\,\,   Q(x,v)< 0
   \end{cases}
\end{equation}
This Finsler's function   is defined and continuous for all $x\in M$ and  for all $v\in T_xM$.  Moreover, it reinstates the initial sign of the quartic. We can define the open conic sets ${\cal A} $ and ${\cal B} $ by the inequalities 
\begin{equation}
    M_{ijkl}(x)v^iv^jv^kv^l> 0\,,\qquad {\rm and} \qquad M_{ijkl}(x)v^iv^jv^kv^l< 0\,,
\end{equation}
respectively. The equation 
\begin{equation}
    M_{ijkl}(x)v^iv^jv^kv^l= 0
\end{equation}
 defines the closed conic set  ${\cal C} $.  It is the first type of singular surface that we meet.

\begin{figure}[h]\label{Fig1}

\centering
\includegraphics[width=0.38\textwidth]{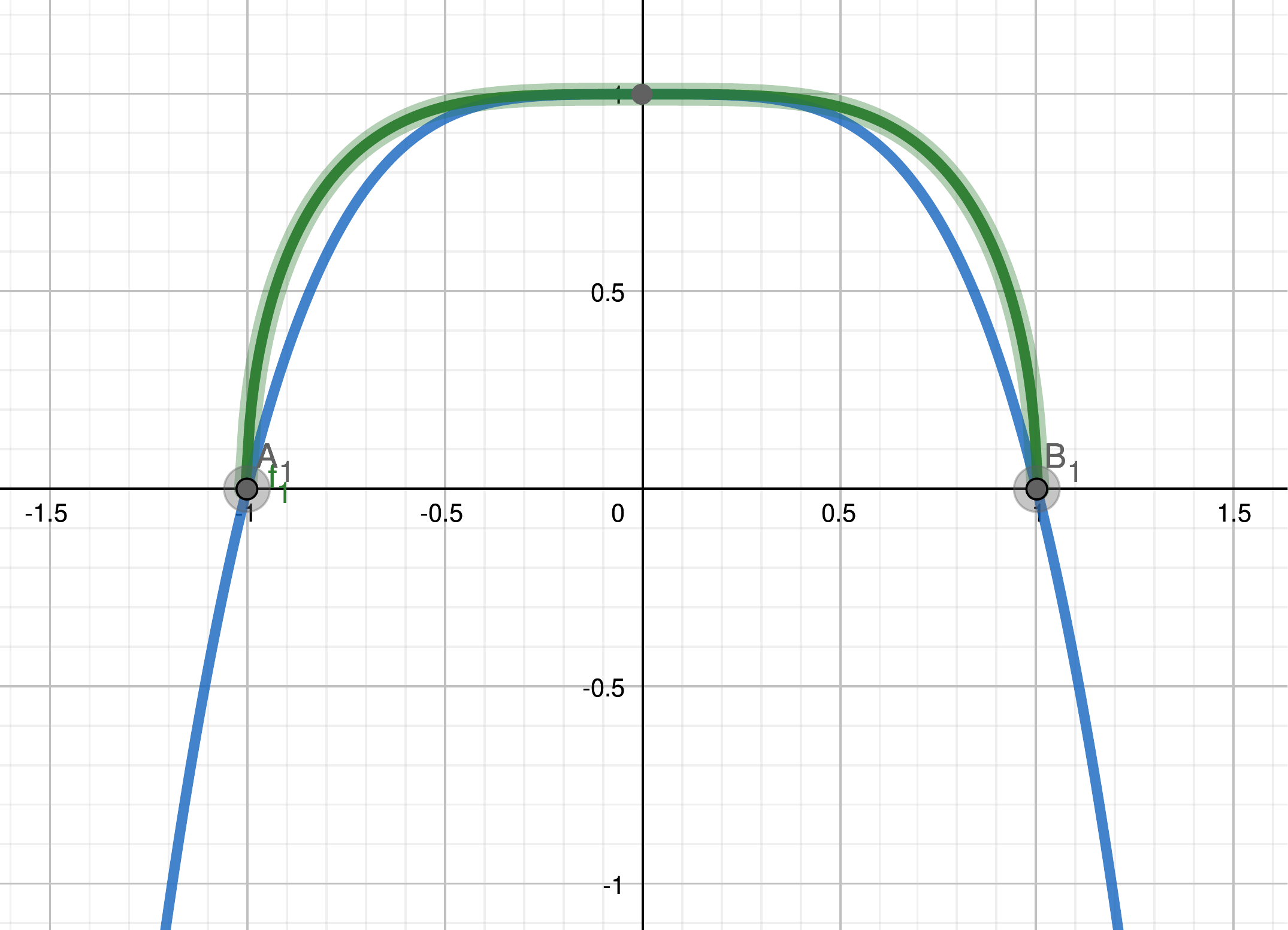}\qquad
\includegraphics[width=0.45\textwidth]{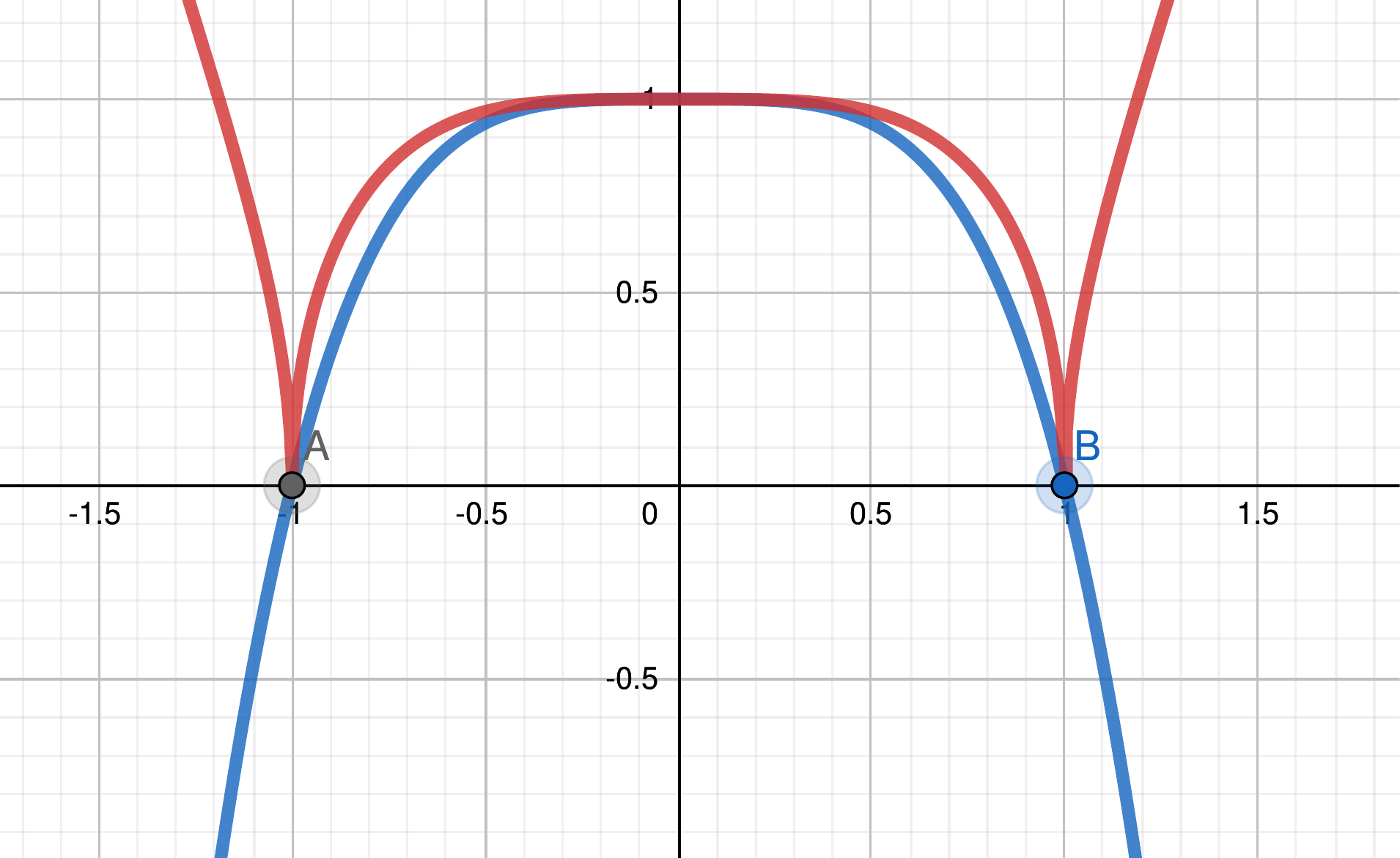}
\includegraphics[width=0.45\textwidth]{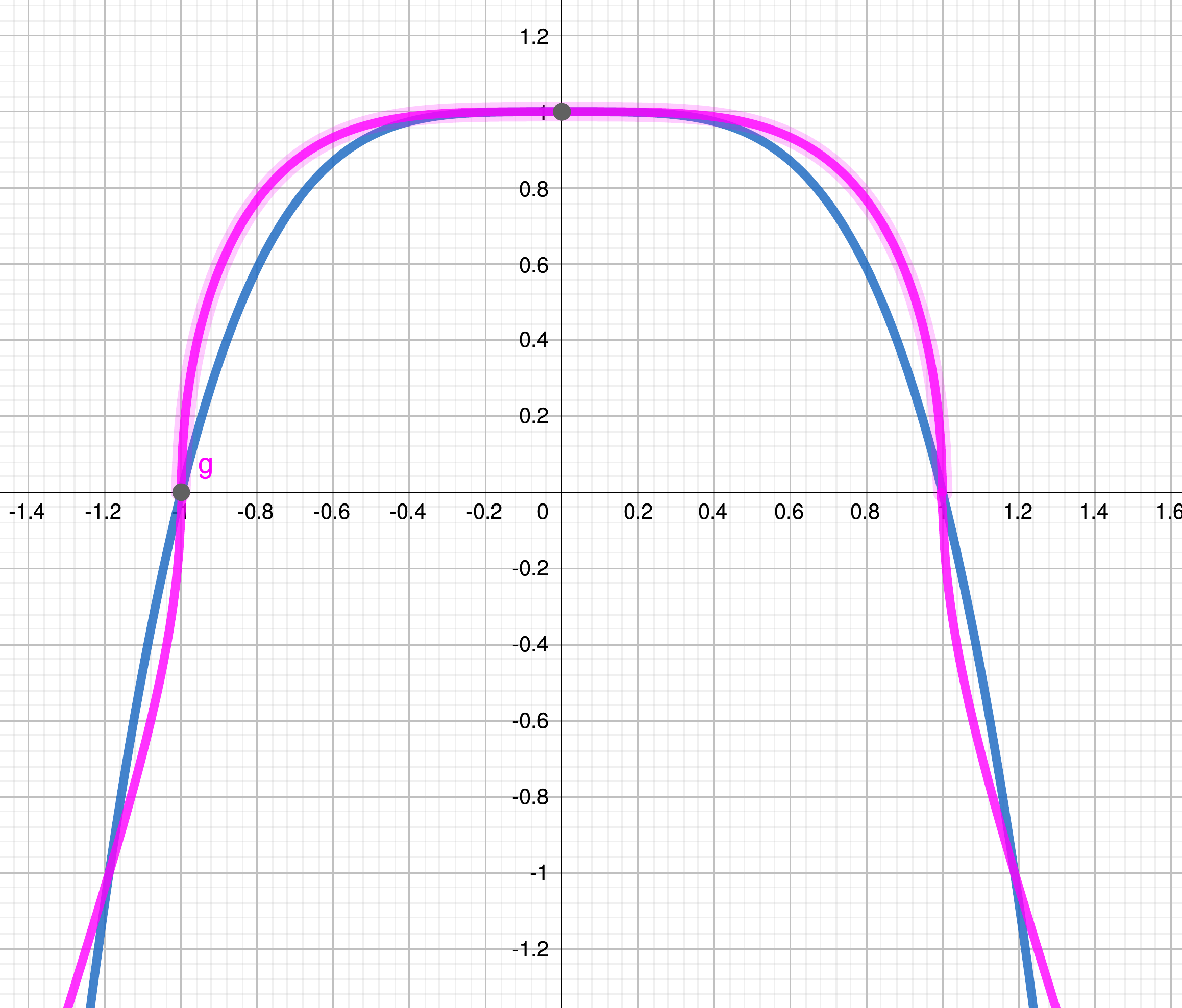}
\caption{Example of different choices of a Finsler function $F(x,v)$ for the same quartic function $Q(x)=1-x^4$ (in blue). Here, $x=\b/\a$ is a direction variable. 
In the first plot, we present the choice  $L(x)=\sqrt{1-x^4}$. In the second plot, we use  $L(x)=\sqrt{|1-x^4|}$. The third plot presents the graph of the 
function $L(x)=\sqrt{|1-x^4|}\,\,\rm{sgn}(Q)$ It has a best matching with the initial quartic.
}
\end{figure}

In Figure \ref{Fig1}, we visualize the difference between three choices (1)--(3) of Finsler's  function $F(x,v)$ with respect to the quartic function $Q(x,v)$. With the third choice we have a best matching   with initial quartic. 
    
\section{Positive definite Riemann's quartics}
In this section we examine Finsler's definition for different examples of Riemann quartic. 
\subsection{Finsler's metric in the Euclidean case}
We start with a positive definite Rieman's quartic $Q(x,v)$. In other words the inequality 
\begin{equation}
    Q(x,v)=M_{ijkl}(x)v^iv^jv^kv^l>0
\end{equation}
holds for  an arbitrary non-zero vector $v$ at an arbitrary point $x\in M$.
In this case, we have a unique choice of the Finsler function 
\begin{equation}
    F(x,v)=\sqrt[4]{Q(x,v)}=\sqrt[4]{M_{ijkl}(x)v^iv^jv^kv^l}.
\end{equation}

The Finsler metric $f_{mn}(x,v)$ is  definite now for all non-zero vectors $v$. 
  With respect to the directional variable, the first order derivative of the function $F(x,v)^2$  takes the form 
 \begin{equation}
     \frac{\partial F(x,v)^2}{\partial v^m}=\frac{Q_{m}(x,v)} {2\sqrt{Q(x,v)}}\,.
 \end{equation}
 Consequently, the Finsler metric is presented as 
 \begin{equation}\label{Q-metric}
     f_{mn}(x,v)=\frac{2QQ_{,mn}-Q_{,m}Q_{,n}}{8\sqrt{|Q|^3}}\,.
 \end{equation}
In term of the tensor $M^{ijkl}$, it is expressed as
\begin{equation}\label{Quart-metric}
    f_{mn}(x,v)=\frac 1{F^6}\left(3F^4 K_{mn} -2N_mN_n\right)\,,
\end{equation} 
where we denote two tensors 
\begin{equation}
K_{mn}= M_{mnij}v^iv^j, \qquad{\mathrm {and}}\qquad      N_m=M_{mijk}v^iv^jv^k\,.
\end{equation}
Observe that this metric is zero-order-homogeneous
\begin{equation}
    f_{mn}(x,\lambda v)=f_{mn}(x,v)
\end{equation}
and satisfies the relation 
\begin{equation}
    f_{mn}(x,v)v^mv^n=F^2(x,v)\,.
\end{equation}
These facts are not enough for the Euclidean  signature of the metric $f_{mn}(x,v)$. 
Positiveness and non-degeneracy of this metric requires
\begin{equation}\label{pos-metric}
    f_{mn}(x,v)u^mu^n> 0
\end{equation}
for an arbitrary non-zero vector  $u^i$, not only for the vector $v^i$. In the two-dimensional case, this requirement is equivalent to the inequality ${\rm det}f_{mn}>0$. In the 4-dimensional case, the determinant inequality is necessary but not sufficient. One has to distinguish between the Euclidean signature $(+1,+1,+1,+1)$ and the mixed-type signature  $(+1,+1,-1,-1)$. 

In fact, it is difficult to satisfy  conditions (\ref{EuclQuartd0}), (\ref{pos-metric}) and the smoothness conditions altogether. In order to emphasize the problems appearing here,  we provide some specific examples.

\subsection{Example 1: Square of a positive-definite quadric}
We start with a square-type of the Riemann quartic
\begin{equation}\label{EuclQuartd1}
  Q(x,v)=\left(g_{ij}v^iv^j\right)^2. 
\end{equation}
Here $g_{ij}$ is a positive definite matrix, then the condition $g_{ij}v^iv^j\ge 0$ holds for all $v\in V$.

Note that  expression (\ref{EuclQuartd1}) can be rewritten as 
\begin{equation}\label{EuclQuartd2}
  Q(x,v)=g_{ij}g_{mn}v^iv^jv^mv^n,  \quad {\mbox {thus} }\quad
  M_{ijkl}=g_{(ij}g_{mn)},
\end{equation}
where the full symmetrization into four indices is applied.
The definition of  Finsler's function in the form 
 \begin{equation}\label{EuclQuartd4}
F(x,v)=\sqrt[4]{Q(x,v)}=\sqrt{g_{ij}v^iv^j}
\end{equation}
 is  allowed and uniquely possible. 
 Due to (\ref{Hess}), the Finsler metric   coincides with the Euclidean one, 
 \begin{equation}\label{EuclQuartd4}
f_{ij}(x,v)=g_{ij}(x)\,.
\end{equation}
The conditions (F1)--(F5) are satisfied. 
 This case is not so interesting because it is pure Riemannian.

\subsection{Example 2: Sum  of the fourth-order powers of the components}
Let us consider the characteristic tensor of the form
\begin{equation}
    M_{ijkl}=\left\{ \begin{array}{ll}
         1 & i=j=k=l;\\
        0 & \mbox{otherwise}.\end{array} \right.
\end{equation}
Consequently the quartic $Q(x,v)$ is presented as a sum of the fourth-order powers of the components of the vector $v$. The first and the second-order derivatives of such expression  evidently vanish on the corresponding axes, see Appendix A. 2 for explicit formulas. Due to Eq.(\ref{Q-metric}), the determinant of  Finsler's metric vanishes on every axis of the vector space separately. 
It means that for a vector $v$ lying on the axis there is another vector $u$ which norm relative to $v$ is zero, $|u|_v=0$.
This problem is well-known, see e.g. \cite{Rund(1959)}. It can be solved by extension of the basic definition of Finsler's structure by the sub-additivity  axiom.

\subsection{Example 3: Product of two positive definite quadrics}
We consider a Riemann quartic 
\begin{equation}\label{EuclProd1}
  Q(x,v)=\left(g_{ij}v^iv^j\right)\left(h_{ij}v^iv^j\right), 
\end{equation}
expanded into  a product of two quadratic forms 
\begin{equation}\label{EuclProd0}
  R(x,v)=g_{ij}v^iv^j,\qquad S(x,v)=h_{ij}v^iv^j.
\end{equation}
where the matrices $g_{ij}$ and $h_{ij}$ are positive definite. 
 Eq.(\ref{EuclProd1}) can be rewritten as 
\begin{equation}\label{EuclProd2}
  Q(x,v)=g_{ij}h_{mn}v^iv^jv^mv^n,  \quad {\mbox {thus} }\quad 
  M_{ijmn}=g_{(ij}h_{mn)}\,.
\end{equation}
The definition of the Finsler function  
\begin{equation}
    F(x,v)=\sqrt[4]{R(x,v) S(x,v)}
\end{equation}
can be considered as a unique possibility. This function is continuous and non-negative for all $v\in T_xM$. The 1-st order homogeneous condition holds for all $v$. The function $F(x,v)$ is smooth for all non-zero $v$. 

Finsler's metric for this expression takes the form
\begin{equation}\label{eq42}
    f_{ij}(x,v)=\frac 1{8\sqrt{R^3S^3}}\left(RS(2SR_{ij}+2RS_{ij})+RS(R_iS_j+R_jS_i)-(R^2S_iS_j+S^2R_iR_j)\right).
\end{equation}
Two last parentheses cancel one another if and only if $R\sim S$.  In this case, Finsler's metric is positive definite. 
In general, it can be positive-definite as well as indefinite. 

Although the function in the right-hand-side of Eq.(\ref{eq42}) is continuous in the whole vector space beyond the origin, it  can  be equal to zero for some non-zero $v$. Moreover it can have positive,  negative or zero determinant.
These facts can be seen from simple 2-dimensional examples.  

In particular, consider a simple quartic   of the form 
\begin{equation}
    Q(x,v)=(\a^2+\b^2)(\a^2+k\b^2)
\end{equation}
with a positive dimensionless parameter $k$, see  Appendix A.3. For sufficiently  large values of the parameter $k$ there is a conic region on the tangent space with the negative determinant of Finsler's metric. In the open competition of this region Finsler's metric is positive-definite, i.e., Euclidean. It degenerates on the boundary between two regions. In other words, Euclidean and Lorentzian signatures of Finsler's metrics are not completely distinct---they can coexist in the same vector space. Certainly, the conditions of the basic definition of Finsler's structure are not satisfied. It means that  these product quartics cannot be embedded into the classical Finsler's space.

\section{Indefinite Lorentzian-type quartics}
We consider Riemann's quartic of the form 
\begin{equation}\label{sect5-1}
    Q(x,v)=M_{ijkl}v^iv^jv^kv^l
\end{equation}
with an {\it indefinite tensor} $M_{ijkl}$. In other words, now $Q(x,v)$ can vanish even for  non-zero vectors $v$. Moreover, for different directions of the  vector $v$,  this function  can obtain  positive and negative values.

\subsection{Beem's Lagrangian vs Finsler's function}
In the indefinite case, it is more convenient to work  with the second order positive homogeneous function $L(x,v)$, called Beem's Lagrangian instead of the 1-st order homogeneous Finsler's function. 
This construction, see \cite{Beem(1970)}, takes into account the fact that Finsler's function  is applied only for determination of the trajectories in  space. In the definition of Finsler's metric, {\it the square of Finsler's function} $F(x,v)^2$ is used instead of  the function $F(x,v)$ itself.  Well-known that even in the ordinary Minkowsky space the vectors can have  positive, negative, and zero squared norms. Consequently, the  square of the Finsler function $F(x,v)^2$ has to be able to give values of different signs. In Beem's construction, the non-negative Euclidean Finsler function $F(x,v)$  is replaced   by a Lagrangian $L(x,v)$ of an arbitrary real value. 
The function $L(x,v)$ is assumed to be definite and continuous at  all points $x\in M$ and for all vectors $v\in T_xM$.  Furthermore, it is assumed to be  differentiable (at least of class $C^4$) for all $x\in M$ and for all nonzero $v\in  T_xM$. The  Finsler metric 
\begin{equation}\label{eq:g}
f_{ij}(x,v)= \frac 12\,\frac{\partial^2 L(x,v)}{\partial v^i\partial v^j}\,.
\end{equation}
 is required to be non-degenerate and of Lorentzian signature {\it for all} $v \neq 0$. 
   
  For indefinite Riemann's quartic, we  use the Lagrangian of the form 
 \begin{equation}
     L(x,v)=\sqrt{|Q(x,v)|}{\rm sgn} (Q)= \sqrt{|M_{ijkl}v^iv^jv^kv^l|}\,{\rm sgn} (Q).
 \end{equation}
  With respect to the directional variable, the gradient of the Lagrangian   takes the form 
 \begin{equation}
     \frac{\partial L(x,v)}{\partial v^m}=\frac{Q_{m}(x,v)} {2\sqrt{|Q(x,v)|}}\,{\rm sgn} (Q)\,.
 \end{equation}
The Finsler metric is presented by the expression
 \begin{equation}\label{Q-metric}
     f_{mn}=\frac{2QQ_{,mn}-Q_{,m}Q_{,n}}{8\sqrt{|Q|^3}}\,{\rm sgn}(Q)\,.
 \end{equation}
 It is clear that derivatives of an arbitrary order exist on the entire tangential space only beside the conic  $Q(x,v)=0$. In order to determine the signature of the metric $f_{mn}$ we need to calculate its  determinant. In four-dimensional case, 
 \begin{equation}\label{Q-metric1}
     {\rm det}(f_{mn})=\frac {{\rm det}\left(2QQ_{,mn}-Q_{,m}Q_{,n}\right)}{8^4Q^6} 
 \end{equation}
 One can guess that for different directions of the vector $v$, this expression can obtain arbitrary real values, i.e., we have here in general a multi-signature regime.  Consider some special examples.

    \subsection{ Square of an indefinite quadric}
    Let  Riemann's quartic be given in the form 
\begin{equation}\label{EuclQuartd5-1}
  Q(x,v)=\left(g_{ij}v^iv^j\right)^2
\end{equation}
 with an {\it arbitrary indefinite tensor} $g_{ij}$. 
 
 In physics literature, the Lagrangian function  is assumed in the form 
  \begin{equation}\label{EuclQuartd5-2}
     L(x,v)={g_{ij}v^iv^j}\,.
 \end{equation}
 This expression is continuous and differentiable for all vectors $v$. 
 Finsler metric in this case takes the form 
 \begin{equation}\label{EuclQuartd5-3}
    f_{ij}=g_{ij}\,.
    \end{equation}
 In Beem's construction, see \cite{Beem(1970)}, the Finsler function  is taken in the form,  
 \begin{equation}
     F(x,v)=\sqrt{g_{ij}v^iv^j}.
 \end{equation}
 This expression is available only in the case $g_{ij}v^iv^j\ge 0,$ i.e., only for some set of {\it ``admissible vectors"}, \cite{Asanov(1985)}. Due to the homogeneity of the Riemann quadric, the  region of these admissible vectors are necessarily  {\it conical}. 
 
 Note that a  {\it conical subset} $U$ of the vector space $V$ is defined as a collection of vectors $v\in U$ such that $\lambda v\in U$ for all positive real $\lambda$.  In particular, the entire space $V$, the half-space, the straight lines going through the origin,  and even the origin $0$ itself are conics. 
 
 The smoothness conditions for the function (\ref{EuclQuartd5-2}) hold only in the open region of the conic $g_{ij}v^iv^j> 0$. These analytic facts demonstrate the advantage in considering the Beem Lagrangian with respect to the Finsler function itself.

 \subsection{Product of a positive-definite and  indefinite quadratics}
 Consider a Riemann quartic presented as a product of two quadrics 
 \begin{equation}
     Q(x,v)=(g_{ij}v^iv^j)(h_{ij}v^iv^j),
 \end{equation}
 where $g_{ij}$ is a positively-definite matrix, while $h_{ij}$ is an indefinite one. 
 We assume a Lagrangian of the form
 \begin{equation}
      L(x,v)=\sqrt{(g_{ij}v^iv^j)|h_{ij}v^iv^j|}\, {\rm sgn} \,Q.
 \end{equation}
 This expression is defined and continuous for all vectors $v$. It is differentiable for all $v$ such that $h_{ij}v^iv^j\ne 0$. In this open conic region, Finsler's metric (\ref{Q-metric}) is defined and smooth. 
 As we can see from the two-dimensional examples (Appendix C), there are two possible cases. 
 
 There  are such pairs of tensors $(g_{ij},h_{ij})$ for which the resulting Finsler's metric is indefinite for some directions of the vector $v$. The singular hypersurface $h_{ij}=0 $ is lying into this area. In the open complement of this Lorentzian area, the metric is positive-definite (Euclidean). On the boundary between two areas, the determinant of the metric vanishes, i.e., the metric is degenerated. 
 
 On the other hand, for some pairs of metrics $(g_{ij},h_{ij}$, the resulting metric is Euclidean, at least for some conic subset of the tangent space.

 \subsection{Product of two indefinite quadratics}
 We consider a Riemann quartic presented as a product of two indefinite quadratics  
 \begin{equation}\label{LL-product}
     Q(x,v)=(g_{ij}v^iv^j)(h_{ij}v^iv^j).
 \end{equation}
 We assume a Lagrangian of the form
 \begin{equation}
      L(x,v)=\sqrt{|(g_{ij}v^iv^j)(h_{ij}v^iv^j)|}\, {\rm sgn} \,Q.
 \end{equation}
 Finsler's metric (\ref{Q-metric}) is defined and smooth, in  open conic region $Q(x,v)\ne 0$, and singular at the hypersurfaces $Q(x,v)=0$. 
 As we can see from the two-dimensional examples (Appendix C), the {\it Finsler metric of a product of two indefinite forms is indefinite for all directions of the vector $v$.} This fact must be important also from the physics point of view, since the case (\ref{LL-product}) is realized in some viable examples from crystal optics, see Sect.8. 
 
 Although we demonstrated this fact only for special diagonal tensors, it can be seen as a general case  (at least in the two-dimensional) since every pair of symmetric matrices can be diagonalized.

 \section{(Pseudo-)Finsler spacetime: a new construction.}
 In this section, we make an attempt to modify the classical  definition of the Finsler structure exhibited above.  Recall that our main aim is to provide a construction that allows us to include  the Riemann quartic.
 As we have already seen, even in the Euclidean case, the standard definition of the Finsler's structure is too restrictive. In general, Finsler's metric is degenerated on  some  hypersurfaces of the entire tangent space.   In the pseudo-Euclidean case, additional singular hypersurfaces where the metric is not defined appear. 
 
 Moreover, as we have seen from the examples above,    the positive-definite and the indefinite structures can coexist on the same tangential vector space. Consequently it does not seem to be productive to separate between Euclidean and Lorentzian signature spaces and to look for different definitions for  them.

 Consequently, we are looking for a weaker definition of Finsler's space that allows us to include various singular subsets. We try to classify these subsets. 
 In fact, one can guess that different singular subsets correspond to different physics phenomena such as wave fronts, caustics, etc.  
 \subsection{Definition}
 \subsubsection{Building block. Continuity} We require the basic Finsler's function  to be defined and continuous over the entire  tangent bundle, i.e. for all points of $M$ and for all vectors of $T_xM$. Since the consideration of the non-negative  first-order Finsler's function is too restrictive for our purposes,  we start with the indefinite second-order Beem's  Lagrangian. So our first condition is 
 \begin{itemize}
     \item[{\bf A}] Let a   positively homogeneous  second-order Lagrangian function  $L(x,v)$ be {\it defined and continuous} over the entire tangent bundle $TM$, i.e. for all  $x\in M$ and for all  $v\in T_xM$. In the entire vector space, we identify three subsets:
     \begin{itemize}
         \item[(i)] The open  conic of {\it spacelike vectors} is the set 
     $${\mathcal A}=\{v\in T_xM\, \mbox{such that }\, L(x,v)>0\,\mbox{for all} \, x\in M\};$$
     \item[(ii)]  The open  conic of {\it timelike vectors} is the set 
     $${\mathcal B}=\{v\in T_xM\, \mbox{such that}\, L(x,v)<0\,\mbox{for all} \, x\in M\};$$
     \item[(iii)]  The open  conic of {\it lightlike (null) vectors} is the set 
     $${\mathcal C}=\{v\in T_xM\, \mbox{such that}\, L(x,v)=0\,\mbox{for all} \, x\in M\}.$$
     \end{itemize}
 \end{itemize}
 \subsubsection{Smoothness} 
 Differentiability of the function  $L(x,v)$ with respect to the position variable $x$ describes the properties of the  manifold  $M$ considered as a differential manifold. Since we are not interested right now in the position singularities of the structure we merely assume the function $L(x,v)$ to be smooth (differential enough time) at almost all points of $M$. Finsler's structure is endowed with another type of differentiability that is considered with respect to the vector variable $v$. 
 
 \begin{itemize}
     \item (B) We require: \\
 $L(x,v)$ to be  differentiable in the open  set  $T_xM\backslash{\mathcal D}_1$;\\
 $L(x,v)$ to be  twice differentiable in the open set  $T_xM\backslash{\mathcal D}_2$;\\
 $\cdots$\\
 $L(x,v)$ to be  $n$-time differentiable in the open  set  $T_xM\backslash{\mathcal D}_n$;
 \end{itemize}
 Due to the homogeneity of $L(x,v)$ the singular sets ${\mathcal D}_i$ are open conics. Moreover, 
 \begin{equation}
     {\mathcal D}_n\subset {\mathcal D}_{n-1}\subset \cdots\subset  {\mathcal D}_1\subset  {\mathcal D}_0\,.
 \end{equation}
 \subsubsection{Metricity}
 The Finsler metric 
 \begin{equation}\label{eq:g}
f_{ij}(x,v)= \frac 12\,\frac{\partial^2 L(x,v)}{\partial v^i\partial v^j}\,.
\end{equation}
 is defined in the open  set $T_xM\backslash{\mathcal D}_2$. 
 \begin{itemize}
     \item (C) We denote:\\
     ${\mathcal E}$ --- The closed conic set where the metric $f_{ij}$ is degenerated; \\
     ${\mathcal F}$ --- The open conic set where the metric $f_{ij}$ is of the Euclidean signature; \\
     ${\mathcal G}$ --- The open conic set where the metric $f_{ij}$ is of the Lorentzian signature; \\
     ${\mathcal H}$ --- The open conic set where the metric $f_{ij}$ is of the mixed signature $(-,-,+,+)$.
 \end{itemize}
 
 \subsection{Two-dimensional examples}
 In this section, we present the characteristic  sets ${\mathcal A},\cdots,{\mathcal H}$ for several two-dimensional examples of Riemann quartic. Explicit calculations are provided in  Appendices. Note some general features related to the homogeneous quartic nature of the Finsler structure. Although,  in general, the characteristic  sets ${\mathcal A},\cdots,{\mathcal H}$ are different, in the case of Riemann's quartic some of them have come to be the same. In particular, 
 \begin{itemize}
 
     \item Due to the analytical properties of the quartic the derivative sets ${\mathcal D}$ are  the same for  for all order, i.e., ${\mathcal D}={\mathcal D}_i$ for all $i$. 
     \item The non-differentiability emerges only on the set of nullity of the Lagrangian, i.e. ${\mathcal C}={\mathcal D}$.
     \item In two-dimensional case, we cannot distinguish between the sets ${\mathcal G}$ and ${\mathcal H}$.
 \end{itemize}  
 In the following table we denote by $V$ the vector space at a  general point without the origin $V=T_xM\backslash\{0\}$. The areas of the positive-definite Finsler's metric are given in blue, while of indefinite metric by yellow. Black lines describe the point where the metric is degenerated. On the red lines, the metric is not defined.

\begin{tabular}{m{4cm} m{5cm} m{5cm}}
\textbf{Quartic} & \textbf{Characteristic Sets} & \textbf{Diagram} \\
  \hline
 $(\a^2+\b^2)^2$ & ${\mathcal A}=V$\par${\mathcal B}=\varnothing$; \par${\mathcal C}=\{0\}$\par${\mathcal D}=\varnothing$;\par ${\mathcal E}=\varnothing$;\par ${\mathcal F}=V$;\par${\mathcal G}=\varnothing$    & \begin{minipage}[t]{2.90\linewidth}
\includegraphics[width=0.42\textwidth]{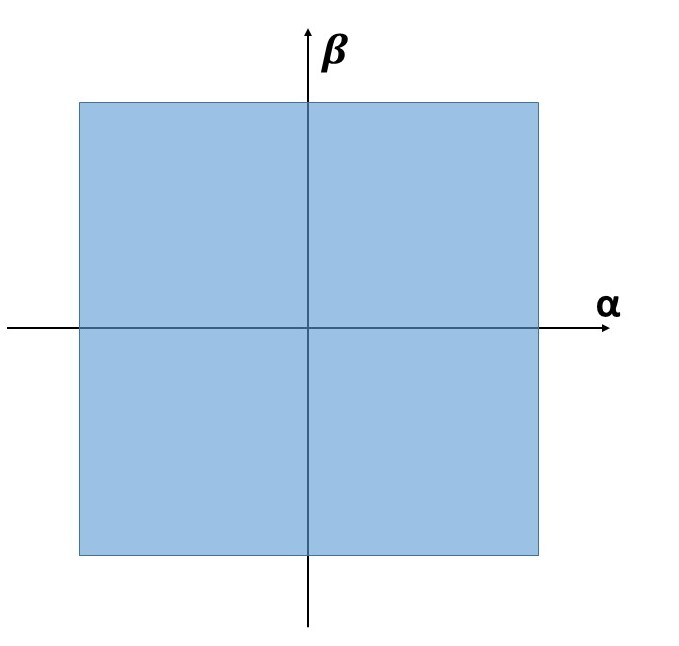}
\end{minipage} \\
 $\a^4+\b^4$ & ${\mathcal A}=V$\par${\mathcal B}=\varnothing$; \par${\mathcal C}=\{0\}$\par${\mathcal D}=V$;\par ${\mathcal E}=\{v\in V|\a\b=0\}$;\par ${\mathcal F}=V\backslash {\mathcal E} ;$\par${\mathcal G}=\varnothing$       & \begin{minipage}[t]{2.90\linewidth}
\includegraphics[width=0.42\textwidth]{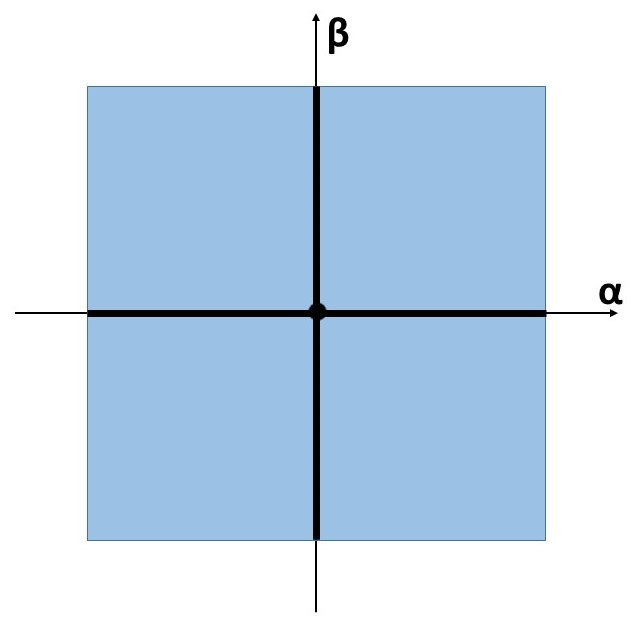}
\end{minipage} \\
 $(\a^2-\b^2)^2$ & ${\mathcal A}=V$\par${\mathcal B}=\varnothing$; \par${\mathcal C}=\{v\in V|\a=\pm\b\}$\par${\mathcal D}={\mathcal C}$;\par ${\mathcal E}=\varnothing$;\par ${\mathcal F}=\varnothing$\par${\mathcal G}=V\backslash {\mathcal C}$      & \begin{minipage}[t]{2.90\linewidth}
\includegraphics[width=0.42\textwidth]{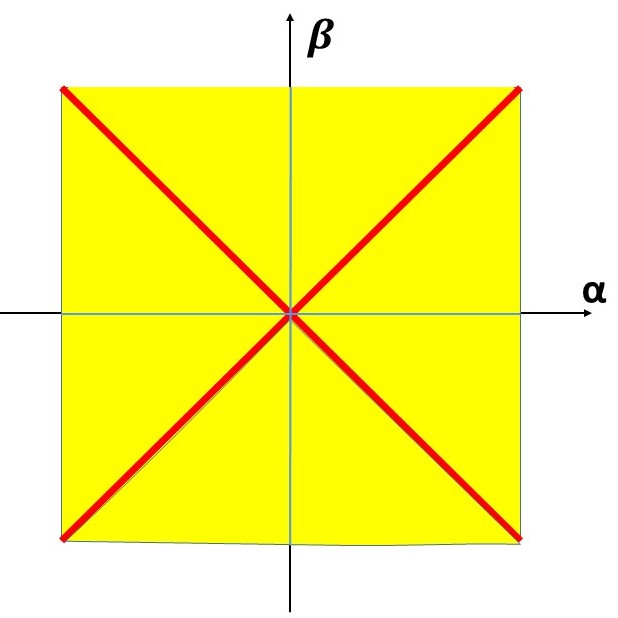}
\end{minipage} \\

\end{tabular}

 \begin{tabular}{m{4cm} m{5cm} m{5cm}}
\textbf{Quartic} & \textbf{Characteristic Sets} & \textbf{Diagram} \\
  \hline
   $\a^4-\b^4$ & ${\mathcal A}=\{v\in V|\a^2>\b^2\}$;\par${\mathcal B}=\{v\in V|\a^2<\b^2\}$; \par${\mathcal C}=\{v\in V|\a^2=\b^2\}$;\par${\mathcal D}=\mathcal C$;\par ${\mathcal E}=\{v\in V|\a\b=0\}$;\par ${\mathcal F}=\varnothing$\par${\mathcal G}=V\backslash ({\mathcal C}\cup{\mathcal E})$     & \begin{minipage}[t]{2.90\linewidth}
\includegraphics[width=0.42\textwidth]{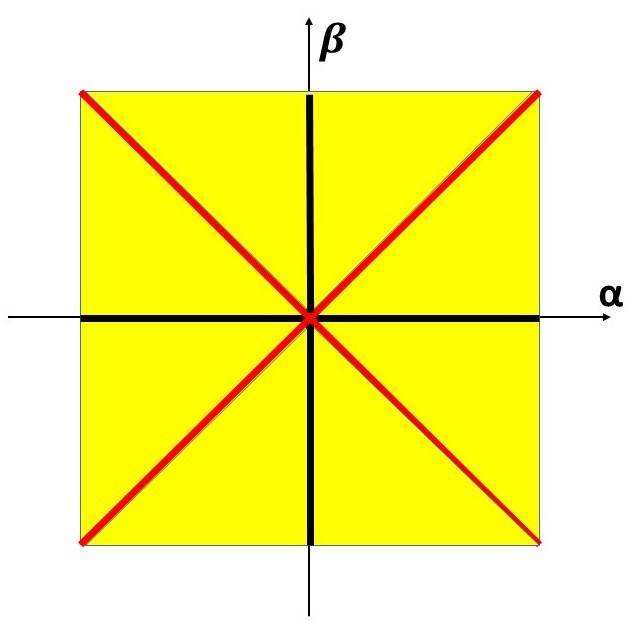}
\end{minipage} \\
 $(\a^2 + \b^2)(\a^2+k\b^2)$ & ${\mathcal A}=V$\par${\mathcal B}=\varnothing$; \par${\mathcal C}=\{0\}$\par${\mathcal D}=\varnothing$;\par ${\mathcal E}\ne\varnothing$;\par ${\mathcal F}\ne\varnothing$;\par${\mathcal G}\ne\varnothing$    & \begin{minipage}[t]{2.90\linewidth}
\includegraphics[width=0.42\textwidth]{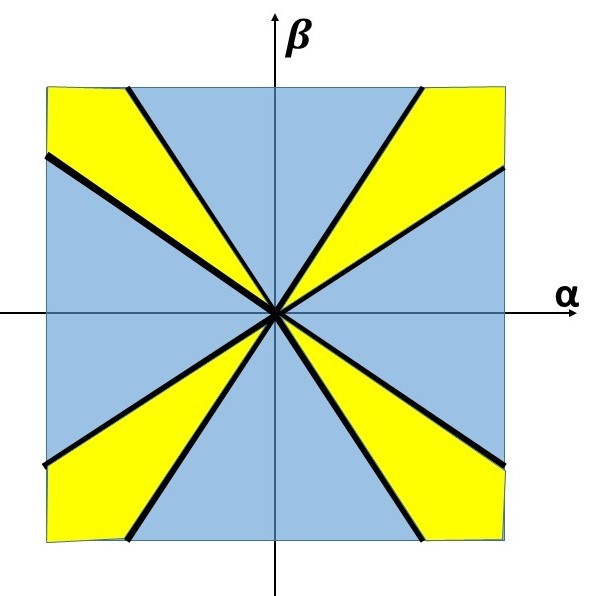}
\end{minipage} \\
 $(\a^2 + \b^2)(\a^2-k\b^2)$ & ${\mathcal A}=\{v\in V|\a^2>k\b^2\}$\par${\mathcal B}=\{v\in V|\a^2<k\b^2\}$; \par${\mathcal C}=\{v\in V|\a^2=k\b^2\}$\par${\mathcal D}={\mathcal C}$;\par ${\mathcal E}\ne\varnothing$;\par ${\mathcal F}\ne\varnothing ;$\par${\mathcal G}\ne\varnothing$       & \begin{minipage}[t]{2.90\linewidth}
\includegraphics[width=0.42\textwidth]{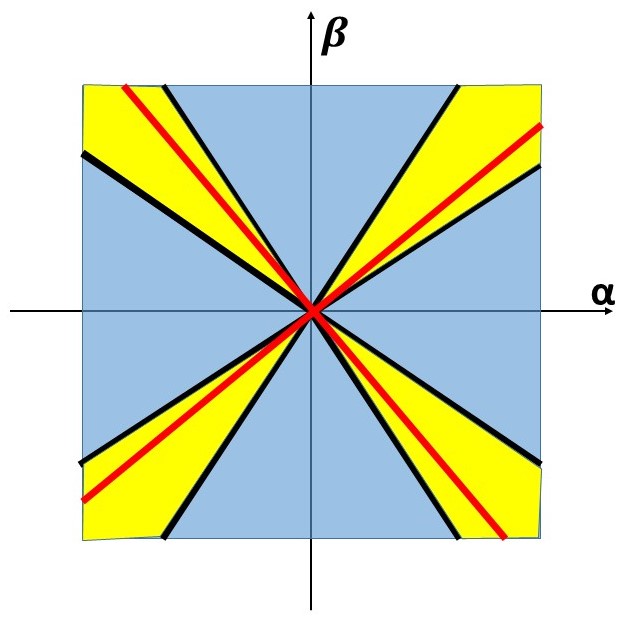}
\end{minipage} \\
 $(\a^2 - \b^2)(\a^2-k\b^2)$ & ${\mathcal A}=V$\par${\mathcal B}=\varnothing$; \par${\mathcal C}=\{v\in V|\a=\pm\b\}$\par${\mathcal D}={\mathcal C}$;\par ${\mathcal E}=\varnothing$;\par ${\mathcal F}=\varnothing$\par${\mathcal G}=V\backslash {\mathcal C}$      & \begin{minipage}[t]{2.90\linewidth}
\includegraphics[width=0.42\textwidth]{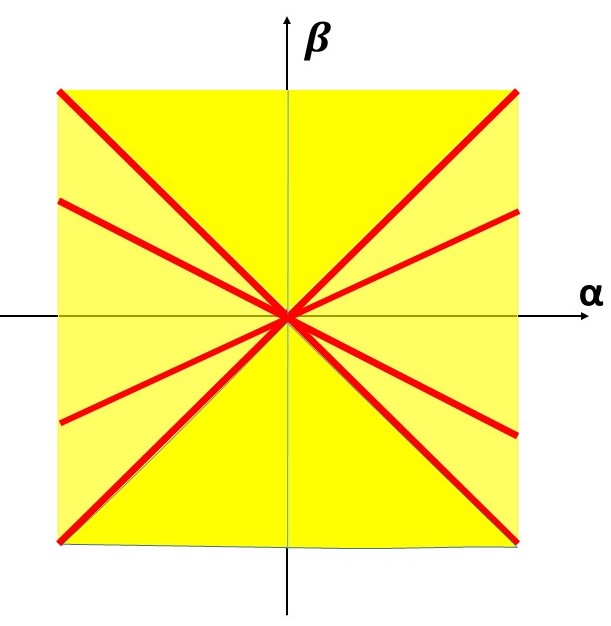}
\end{minipage}
\end{tabular}

 \subsection{Electromagnetic wave propagation in uniaxial crystal}
 As a viable illustration of the Finsler construction in physics, we consider a quartic appearing in  electromagnetic wave propagation in anisotropic media. As we discussed in Sect.1, a general dispersion relation is described by a  fourth order polynomial. A special case of a media characterized by two anisotropic  tensors $\varepsilon^{\a\b}$ and $\pi_{\a\b}$, as it is presented in \cite{Itin:2009fj}, is rather complicated for Finsler's analysis. Thus we consider a simplest anisotropic media appearing in uniaxial crystal. In this case, the impermeability  matrix $\mu_{\a\b}$ is assumed to be isotropic, while the permittivity  matrix $\varepsilon^{\a\b}$ has only two independent components
 \begin{equation}
     \mu_{\a\b}=\mu g_{\a\b}\qquad \varepsilon^{\a\b}={\rm diag}(\varepsilon_o,\varepsilon_o,\varepsilon_e)
 \end{equation}
 Here the Greek indices change in range $\a,\b,\cdots=1,2,3$. The optic axis is assumed to coincide with the $z$-axes. The value $\varepsilon_o$ corresponds to the {\it ordinary wave}, while $\varepsilon_e$ is used for the {\it extraordinary wave}. For the wave vector $q$ parametrized as $q=(\omega,q_1,q_2,q_3)$ the dispersion relation reads, \cite{Landau}
 \begin{equation}
    Q(\omega,q_\a)= (\varepsilon_o\mu\omega^2-q_1^2-q_2^2-q_3^2)(\mu\varepsilon_o\varepsilon_e\omega^2-\varepsilon_o(q_1^2+q_2^2)-\varepsilon_eq_3^2)=0
 \end{equation}
 Instead of presenting a rather involved expression for the fourth-order Finsler's metric corresponding to $Q(\omega,q_\a)$, we analyze Finsler's structure on different second-order cross-sections.
 
 \vspace{0.3cm}
 
{\it (i) Cross-section $q_1=q_2=0$.}

\noindent In this case,
 \begin{equation}
    Q(\omega,q_3)= \varepsilon_e(\varepsilon_o\mu\omega^2-q_3^2)^2\,.
 \end{equation}
Consequently, Finsler's metric is given as in Appendix B.1 by 
\begin{equation}
    f_{ij}(v)=\varepsilon_e{\rm diag}(\varepsilon_o\mu, -1).
\end{equation}
 For $\varepsilon_o\mu>0$, this indefinite (Lorentzian) expression describes  an ordinary wave with the velocity $v=\sqrt{\varepsilon_o\mu}$.
 
 \vspace{0.3cm}
 
{\it (ii) Cross-section $q_2=q_3=0$}
 
 \noindent This case corresponds to the example of a product of two indefinite quadratic expressions, Appendix C.3. The quartic is given by
 \begin{equation}
     Q(\omega,q_1)=\varepsilon_o(\varepsilon_o\mu\omega^2-q_1^2)(\varepsilon_e\mu\omega^2-q_1^2).
 \end{equation}
 Finsler's metric takes the form given in Eq.(\ref{LL-metric}) with the parameters
 \begin{equation}
     \a=q_1,\quad \b=\omega\sqrt{\varepsilon_o\mu},\quad k=\frac{\varepsilon_e}{\varepsilon_o}\,.
 \end{equation}
 The determinant of this metric is given in Eq.(\ref{LL-det}).
It is strictly negative and degenerates at the surfaces $\varepsilon_o\mu\omega^2-q_1^2=0$ --- ordinary wave and $\varepsilon_e\mu\omega^2-q_1^2=0$ --- extraordinary wave.
Thus Finsler's metric is indefinite (Lorentzian) in all directions of the vector $v$. 

 \vspace{0.3cm}
 
{\it (iii) Cross-section $\omega=q_3=0$}\\
The quartic is given by 
\begin{equation}
     Q(q_1,q_2)=\varepsilon_o(q_1^2+q_2^2)^2.
 \end{equation}
 This case (Appendix A1) corresponds to the standard Euclidean metric 
 \begin{equation}
     f_{ij}(v)=\varepsilon_o\,{\rm diag}(1, 1).
 \end{equation}
 The unit spheres are the standard circles. Physically this cross-section corresponds to the ordinary wave. 

 \vspace{0.3cm}
 
{\it (iv) Cross-section $\omega=q_2=0$}\\
The quartic is given by 
\begin{equation}
     Q(q_1,q_3)=(q_1^2+q_2^3)\left(\varepsilon_o\, q_1^2+{\varepsilon_e}\,q_2^3\right).
 \end{equation}
 With the parameters 
 \begin{equation}
     \a=q_1,\quad \b=q_2,\quad k=\frac{\varepsilon_e}{\varepsilon_o}
 \end{equation}
 Finsler's metric is described  by Eq.(\ref{EE-metric}). The determinant is given in Eq.(\ref{EE-det}) of Appendix C1. 
 Note that for sufficiently large values of the parameter $k={\varepsilon_e}/{\varepsilon_o}$ there is an area in which the signature of Finsler's metric is modified. This fact indicates the complicated global features of Finsler's metric.

 \section{Conclusion}
 In this paper, we consider Riemann's quartic expression that historically was proposed as a first example of Finsler's structure. Quite surprisingly, this pure mathematical example of Riemann turns out to be the proper description of the electromagnetic wave propagation phenomena in linear media. We demonstrate that in most cases Riemann's quartic cannot be embedded in the strict classical definition of Finsler's geometry. In most cases, this definition is broken-down on various hypersurfaces of the tangent vector space. Moreover, we indicate a possibility of coexistence of areas with different signatures on the same tangent space. These areas are separated by singular conic hypersurfaces  with the degenerated Finsler's metric. As a result, most examples of Riemann's quartics turn out to be non-Finslerian. 
 
 Our proposal is to consider a {\it weakened Finsler structure}  that allows to include much more physically meaningful  examples, in particular Riemann's quartics of an arbitrary sign. A classification of singular surfaces is worked out.  Our construction is not restricted to a special signature of the space. The future line of consideration is to study the relations between the singular sets of Finsler's structure and observable phenomena of electromagnetic wave propagation. 
 \appendix

\section{Two-dimensional positive definite  quartics}
In order to clarify the conditions presented in the  definition of Finsler's structure, we consider some explicit 2-dimensional examples.    We use Beem's construction with the second-order homogeneous Lagrangian instead of the first-order homogeneous Finsler's function itself. The  dependence of a point $x$ in the Lagrangian $L(x,v)$ is suppressed while the direction variable is presented in the form $v=(v^1,v^2)=(\a,\b)$. Note that the 2-dimensional case has some special similarity to the 4-dimensional one. In both cases, a metrics $f_{ij}$ of the  Euclidean  signature have to satisfy the condition
\begin{equation}\label{Eucl-cond}
    {\rm det}f_{ij}>0.
\end{equation}
In the 2-dimensional case, this condition is necessary and sufficient since it extracts the metrics of the signatures $(+1,+1)$ and $(-1,-1)$, so that both can  be considered as Euclidean. 
In the 4-dimensional case, (\ref{Eucl-cond}) is only necessary since it is satisfied  by  the metrics of the signature $(+1,+1,+1,+1)$ and  $(-1,-1,+1,+1)$. 
In this case, we need additional algebraic   inequalities in order to extract the Euclidean signature. 
For the Lorentzian signature, the condition 
\begin{equation}\label{Lor-cond}
    {\rm det}f_{ij}<0 
\end{equation}
is necessary and sufficient in two-dimensional and  four-dimensional cases.

\subsection{Square of Euclidean quadric}
The simplest positive definite quartic expression is presented as the square of the Euclidean quadratic
\begin{equation}\label{ap1-1}
    Q(v)=(\a^2+\b^2)^2\,.
\end{equation}
The unique possible Lagrangian for this expression is given by
\begin{equation}\label{ap1-2}
    L(v)=\a^2+\b^2\,.
\end{equation}
This expression is positive for all non-zero vectors $v$ and vanishes only at the origin. Moreover, it is continuous and smooth for all $v\in V$. The corresponding Finsler metric is Riemannian 
\begin{equation}
    f_{ij}={\rm diag}(1,1)
\end{equation}
and well-definite for all vectors in the plan. Finsler's function 
\begin{equation}
    F(v)=\sqrt{\a^2+\b^2}
\end{equation}
is well-definite for all $v$ and smooth everywhere except the origin $v=(\a,\b)=0$.  

\subsection{One more positive defined quartic}
As an example of a more general Riemann's quartic of the Euclidean type, we consider the expression  of the form 
\begin{equation}\label{L1}
    Q(v)={\a^4 + \b^4}.
\end{equation}
The unique possible Lagrangian for this quartic is given by 
\begin{equation}\label{L1}
    L(v)=L(\a,\b)=\sqrt{\a^4 + \b^4}\,.
\end{equation}
This expression is well-definite and continuous for all vectors $v$ and has  the required second order of homogeneity. Note that the expression (\ref{L1})  can be regarded as a special case of a quartic  decomposable into a product of two positive definite quadratic factors
\begin{equation}\label{L1x}
    L(v)=L(\a,\b)=\sqrt{(\a^2 -\sqrt{2}\a\b+ \b^2)(\a^2 +\sqrt{2}\a\b+ \b^2)}\,.
\end{equation}
The vector of the partial derivatives of the Lagrangian (\ref{L1}) 
\begin{equation}
    L_{,i}=\frac {2}{\sqrt{\a^4 + \b^4}} 
    \begin{pmatrix}
\a^3\\
\b^3
\end{pmatrix}\,
\end{equation}
is continuous and non-zero for all $v\ne 0$. In the origin $v=(\a,\b)=0$, the function $L(v)$ is non-differentiable. 
Calculating the Hessian of $L(v)$, we obtain Finsler's metric in the form 
\begin{equation}\label{A2-8}
    f_{ij}=\frac {1}{(\a^4 + \b^4)^{3/2}}
    \begin{pmatrix}
\a^2(\a^4 + 3\b^4) & -2\a^3\b^3\\
-2\a^3\b^3 & (3\a^4 + \b^4)\b^2
\end{pmatrix}.
\end{equation}
The determinant of this matrix 
   \begin{equation}\label{A-det}
       {\rm det}(f_{ij})=\frac{3\a^2\b^2}{\a^4 + \b^4}
   \end{equation}
is non-negative for all $v$. It is singular only at the origin $v=(\a,\b)=0$. Moreover it vanishes  on two separate axes $\a=0$ and $\b=0$, i.e., for the vectors $v=(0,\b)$ and $v=(\a,0)$, respectively.   Beyond these axes, the metric tensor is positive definite. 
In particular, the norm of an arbitrary nonzero vector $u=(x,y)$ with respect to the metric (\ref{A2-8}) reads
\begin{equation}
    ||u||^2_v=\frac  {1}{(\a^4 + \b^4)^{3/2}}\left(\a^2(\a^4 + 3\b^4)x^2-4\a^3\b^3xy+(3\a^4 + \b^4)\b^2y^2\right)\,,
\end{equation}
or, equivalently,
\begin{equation}
    ||u||^2_v=\frac  {1}{(\a^4 + \b^4)^{3/2}}
    \left((\a^3x + \b^3y)^2+3\a^2\b^2(\b x-\a y)^2\right)\,.
\end{equation}
With respect to the new coordinates $(x,y)\to(\xi,\eta)$
\begin{equation}\label{ap-trans}
    \xi=\frac  {\a^3x + \b^3y}{(\a^4 + \b^4)^{3/4}},\qquad \eta=\frac  {\sqrt{3}\a\b(\b x-\a y)}{(\a^4 + \b^4)^{3/4}}\,
\end{equation}
 the norm expression is given in the standard Euclidean form 
\begin{equation}\label{Eucl-norm3}
    ||u||^2_v=\xi^2+\eta^2\,.
\end{equation}
In order to have in (\ref{ap-trans}) a reversible transformation of coordinates we have to require
\begin{equation}
 \frac  {1}{(\a^4 + \b^4)^{3/2}} \,{\rm det}  \begin{pmatrix}
\a^3 & \b^3\\
\sqrt{3}\a\b^2 & -\sqrt{3}\a^2\b
\end{pmatrix}=-\frac{\sqrt{3}\a\b}{\sqrt{\a^4 + \b^4}}\ne 0.
\end{equation}
It means $\a\b\ne 0$ that is in  correspondence with the expression for the determinant (\ref{A-det}). 
The  expression (\ref{Eucl-norm3}) is positive beyond the axes. However on $\a$-axis, i.e., for $\b=0$, it takes the value $||u||^2_v=x^2$ that is zero for nonzero vectors of the form $u=(0,y)$. A similar singular behavior emerges also in the direction based on  the basis-element vectors $v=(0,\b)$. 

Consequently,  Finsler's metric defined for the Lagrangian (\ref{L1}) has two peculiar  directions in which the geometry degenerates and cannot be considered as pure  Finslerian.

\section{Two-dimensional indefinite quartics}
In this section we consider several  two-dimensional examples of the indefinite quartic 
\begin{equation}
    Q(x,v)=M_{ijkl}v^iv^jv^kv^l.
\end{equation}
In this case, the indices change in the range $i,j,\cdots=1,2$. We use the presentation $v^i=(\a,\b)$. The  quartic $Q(x,v)$ is indefinite--it can have positive, negative, or zero value for different directions of the vector $v$. 

  In the 2-dimensional case (as well as the 4-dimensional case)  the metrics $f_{ij}$ of Lorentzian signature is distinguished by an unique condition
\begin{equation}
    {\rm det}f_{ij}<0.
\end{equation}
In other dimensions, we need more algebraic   inequalities. 
Having in the mind 4-dimensional Lorentz signature of the form $(-1,+1,+1,+1)$ we refer to the vectors of positive/negative Lagrangians (squared Finsler's norms)  as {\it spacelike} and {\it timelike},   respectively. The {\it null vectors} are vectors with a zero Lagrangian.

\subsection{Square of Lorentzian quadric}
We start with a simplest  quartic that is presented as a square of an indefinite quadratic
\begin{equation}\label{ap2-1}
    Q(v)=(\a^2-\b^2)^2\,.
\end{equation}
This expression is positive but vanishes for non-zero vectors of the form $v=(\a,\pm \b)$.

Define the Lagrangian in the standard (special relativistic) form 
\begin{equation}
    L(v)=\a^2-\b^2\,.
\end{equation}
This function  is smooth for all vectors $v$. The Finsler metric has the standard Lorentz form 
\begin{equation}
    f_{ij}(v)={\rm diag}(1,-1)\,.
\end{equation}

Note that three characteristic sets can be determined
\begin{eqnarray}
    {\cal A}&=&\{v\in T_xM\big||\a|>|\b|\} \,\,\Longrightarrow\,\, {\rm spacelike\,\, vectors}\\
    {\cal B}&=&\{v\in T_xM\big||\a|<|\b|\} \,\,\Longrightarrow\,\, {\rm timelike\,\, vectors}\\
    {\cal C}&=&\{v\in T_xM\big||\a|=|\b|\} \,\,\Longrightarrow\,\, {\rm null\,\, vectors}
\end{eqnarray}

Consider different possible definitions of the Finsler function  (norm). In fact they are  not related to the specific choice of the Lagrangian.

(1) Finsler's function of the form 
\begin{equation}
    F(v)=\sqrt{\a^2-\b^2}
\end{equation}
is defined only in the region satisfying $|\a|\ge |\b|$ and differentiable in the open conic $|\a|> |\b|$. 

(2) Another possibility is to define the Finsler function in the absolute value form
\begin{equation}
    F(v)=\sqrt{|\a^2-\b^2|}\,.
\end{equation}
Recall that this construction was proposed by Beem, see (\ref{}). This function is defined for all vectors $v$ and differential besides the set ${\cal C}$. It cannot, however, separate the time-like and space-like vectors. 

(3) The third definition is based on the proposal in Sect. 5. 
\begin{equation}
    F(v)=\sqrt{|\a^2-\b^2|}\,{\rm sgn} L(v)=\left\{ \begin{array}{ll}
        \,\,\, \,\sqrt{\a^2-\b^2}\qquad {\rm for} \qquad |\a|\ge|\b|;\\
        -\sqrt{\b^2-\a^2}\qquad {\rm for}  \qquad |\a|<|\b|.\end{array} 
        \right.
\end{equation}
In this case, the norm of the  space-like vectors is  positive  and of the time-like vectors is negative. The 
norm of the null-vectors is zero.

\subsection{Lorentzian non-birefringence space-time}
We consider now a Lorentzian-type quartic of the form
\begin{equation}
    Q(v)=\a^4 - \b^4\,.
\end{equation}
This expression is well-definite and smooth for all vectors $v=(\a,\b)$. It can be expressed as a product of two quadrics--one Euclidean and one Lorentzian
\begin{equation}
    Q(v)=(\a^2 + \b^2)(\a^2-\b^2)\,.
\end{equation}
Our first task is to construct a proper Lagrangian.
Let us consider different options:

\vspace{0.2cm}
(1) Define the Lagrangian of the form
\begin{equation}
    L(v)=L(\a,\b)=\sqrt{\a^4 - \b^4}=\sqrt{(\a^2 - \b^2)(\a^2+\b^2)}
\end{equation}
This function is definite and continuous only in the region $|\a|\ge |\b|$ and smooth in its interior $|\a|> |\b|$. Then,  the characteristic sets are 
\begin{eqnarray}
    {\cal A}&=&\{v\in T_xM\big||\a|>|\b|\} \,\,\Longrightarrow\,\, {\rm spacelike\,\, vectors}\\
    {\cal B}&=&\emptyset \,\,\Longrightarrow\,\, {\rm timelike\,\, vectors}\\
    {\cal C}&=&\{v\in T_xM\big||\a|=|\b|\} \,\,\Longrightarrow\,\, {\rm null\,\, vectors}
\end{eqnarray}
The gradient of this function is given by 
\begin{equation}
      L_{,i}=\frac {2}{\sqrt{\a^4 - \b^4}} 
      \begin{pmatrix}
\a^3\\
-\b^3
\end{pmatrix}.
\end{equation}
This expression is continuous in the region ${\cal A}$. The same is true for all higher derivatives of $L(v)$. Consequently for all positive integers $i$, the sets of differentiability coincide with the set of positive values of $L(v)$, i.e.,   ${\cal D}_i={\cal A}$ . 
The Hessian yields Finsler's metric of the form 
\begin{equation}
    f_{ij}=\frac {1}{(\a^4 - \b^4)^{3/2}}
    \begin{pmatrix}
\a^2(\a^4 - 3\b^4) & 2\a^3\b^3\\
2\a^3\b^3 & \b^2(\b^4-3\a^4)
\end{pmatrix}.
\end{equation}
 The determinant of this metric, 
   \begin{equation}
       f=\frac{-3\a^2\b^2}{\a^4 - \b^4}\,,
   \end{equation}
is non-positive in the set ${\cal A}$. It is degenerate on the strength line $\b=0$ (the line $\a=0$ is beyond the set ${\cal A}$) . It means that the set of degeneration is defined as  ${\cal E}=\{v\in T_xM|\b= 0\}$. Besides this singular set, the metric is Lorentzian, i.e., ${\cal G}=\{v\in T_xM|\b\ne 0\}$.

 For the norm of an arbitrary vector $u=(x,y)$, we have 
 \begin{equation}
    ||u||^2_v=\frac  {1}{(\a^4 - \b^4)^{3/2}} \left(\a^2(\a^4 - 3\b^4)x^2+4\a^3\b^3xy+
    \b^2(\b^4-3\a^4 )y^2\right)\,,
\end{equation}
   or, equivalently,
    \begin{equation}
    ||u||^2_v=\frac  {1}{(\a^4 - \b^4)^{3/2}} 
    \left((\a^3x + \b^3y)^2 -3\a^2\b^2(\b x-\a y)^2 \right)\,,
\end{equation}
Let us define new coordinates $(x,y)\to(\xi,\eta)$
\begin{equation}\label{trans}
    \xi=\frac{\a^3x + \b^3y}{(\a^4 - \b^4)^{3/4}},\qquad \eta=\frac{\sqrt{3}\a\b(\b x-\a y)}{(\a^4 - \b^4)^{3/4}}\,.
\end{equation}
Then the norm of the vector $u$ is given in the standard  Lorentzian form
\begin{equation}
    ||u||^2_v=\xi^2-\eta^2\,.
\end{equation}
Note that the transformation of the coordinates is applicable only in the regions where it is revertible. So we have to require
\begin{equation}
  {\rm det}  \begin{pmatrix}
\a^3 & \b^3\\
\sqrt{3}\a\b^2 & -\sqrt{3}\a^2\b
\end{pmatrix}=-\sqrt{3}{\a\b}(\a^4+\b^4)\ne 0\,.
\end{equation}
It means that the transformations are available only for $\a\b\ne 0$.

\vspace{0.2cm}
(2) Let us consider a Lagrangian of the absolute value form
\begin{equation}
    L(v)=L(\a,\b)=\sqrt{|\a^4 - \b^4|}=\sqrt{|\a^2 - \b^2|(\a^2+\b^2)}
\end{equation}
The function is non-negative in the whole plane and strictly positive in open conic regions beside the strength lines $|\a|=|\b|$. 
Then,  the characteristic sets are 
\begin{eqnarray}
    {\cal A}&=&\{v\in T_xM\big||\a|\ne|\b|\} \,\,\Longrightarrow\,\, {\rm spacelike\,\, vectors}\\
    {\cal B}&=&\emptyset \,\,\Longrightarrow\,\, {\rm timelike\,\, vectors}\\
    {\cal C}&=&\{v\in T_xM\big||\a|=|\b|\} \,\,\Longrightarrow\,\, {\rm null\,\, vectors.}
\end{eqnarray}

In the regions $|\a|>|\b|$ and $|\a|<|\b|$, the gradient of this function is given by 
\begin{equation}
      L_{,i}=\frac {2}{\sqrt{\a^4 - \b^4}} 
      \begin{pmatrix}
\a^3\\
-\b^3
\end{pmatrix}, \qquad {\rm and}\qquad 
    L_{,i}=\frac {2}{\sqrt{-\a^4 +\b^4}} 
      \begin{pmatrix}
-\a^3\\
\b^3
\end{pmatrix},
\end{equation}
respectively. 
The Hessian yields Finsler's metric of the form 
\begin{equation}
    f_{ij}=\frac {1}{|\a^4 - \b^4|^{3/2}}
    \begin{pmatrix}
\a^2(\a^4 - 3\b^4) & 2\a^3\b^3\\
2\a^3\b^3 & \b^2(\b^4-3\a^4)
\end{pmatrix}.
\end{equation}
 The determinant of this metric reads
   \begin{equation}
       f=\frac{-3\a^2\b^2}{|\a^4 - \b^4|}
   \end{equation}
   Consequently for $\a,\b\ne 0$ the determinant is negative. In a 2-dimensional space (and in a  4-dimensional space, as well) this requirement is sufficient  to deduce that the metric is Lorentzian. 
For this Lagrangian, the norm of an arbitrary vector can be also written in a pure Lorentzian form  $||u||^2_v=\xi^2-\eta^2$. This form is derived by the transformations (\ref{trans}) with an absolute value function added into the denominator. 

\vspace{0.2cm}
(3) Let us consider a Lagrangian of the form
\begin{equation}
    L(v)=L(\a,\b)=\sqrt{|\a^4 - \b^4|}\,{\rm sgn}(Q)=\left\{ \begin{array}{ll}
        \,\,\, \,\sqrt{\a^4-\b^4}\qquad \,\,\,\,\,{\rm for} \qquad |\a|\ge|\b|;\\
        -\sqrt{-\a^4+\b^4}\qquad {\rm for}  \qquad |\a|<|\b|.\end{array} 
        \right.
\end{equation}
The function is definite and continuous  in the whole plane. 
The characteristic sets are 
\begin{eqnarray}
    {\cal A}&=&\{v\in T_xM\big||\a|>|\b|\} \,\,\Longrightarrow\,\, {\rm spacelike\,\, vectors}\\
    {\cal B}&=&\{v\in T_xM\big||\a|<|\b|\} \,\,\Longrightarrow\,\, {\rm timelike\,\, vectors}\\
    {\cal C}&=&\{v\in T_xM\big||\a|=|\b|\} \,\,\Longrightarrow\,\, {\rm null\,\, vectors.}
\end{eqnarray}
and strictly positive in open conic regions beside the strength lines $|\a|=|\b|$.
In the regions $|\a|>|\b|$ and $|\a|<|\b|$, the gradient of this function is given by 
\begin{equation}
      L_{,i}=\frac {2}{\sqrt{\a^4 - \b^4}} 
      \begin{pmatrix}
\a^3\\
-\b^3
\end{pmatrix}, \qquad {\rm and}\qquad 
    L_{,i}=\frac {2}{\sqrt{-\a^4 +\b^4}} 
      \begin{pmatrix}
\a^3\\
-\b^3
\end{pmatrix},
\end{equation}
respectively. 
The Hessian yields Finsler's metric of the form 
\begin{equation}
    f_{ij}=\frac {1}{|\a^4 - \b^4|^{3/2}}
    \begin{pmatrix}
\a^2(\a^4 - 3\b^4) & 2\a^3\b^3\\
2\a^3\b^3 & \b^2(\b^4-3\a^4)
\end{pmatrix}{\rm sgn}(Q).
\end{equation}

 The determinant of this metric 
   \begin{equation}
       f=\frac{-3\a^2\b^2}{|\a^4 - \b^4|}
   \end{equation}
is definite and negative almost everywhere. It vanishes only on the lines $|\a|=0$ and $|\b|=0$ and degenerates on the lines $|\a|=|\b|$. 
In this metric the squared norm of a vector can be also transformed into the standard Lorentzian form $||u||^2_v=\xi^2-\eta^2$.

 \section{Quartic as a product  of two quadratics}
Let us consider an example of a quartic that is expanded into  a product of two  factors 
\begin{equation}
    Q(v)=Q(\a,\b)=\left(g_{ij}v^iv^j\right)\left(h_{ij}v^iv^j\right)
\end{equation}
with two  matrices $g_{ij}$ and $h_{ij}$. Since these two matrices can be transformed simultaneously into the diagonal form, we come to expressions of three possible type
\subsection{ Euclid $\times$ Euclid}
Let us consider a quartic  expression
\begin{equation}
    Q(v)=Q(\a,\b)=(\a^2 + \b^2)(\a^2+k\b^2)\,,
\end{equation}
where $k$ is a dimensionless {\it positive} numerical parameter. 
Thus we have a positive-definite quartic. 
For this quartic function, the unique possible Lagrangian is given by 
\begin{equation}
    L(v)=L(\a,\b)=\sqrt{(\a^2 + \b^2)(\a^2+k\b^2)}\,.
    \end{equation}
This function is definite and continuous for all vectors $v=(\a,\b)$.     
    The gradient of $L(v)$  takes the form
\begin{equation}
      L_{,i}=\frac {1}{\sqrt{(\a^2 + \b^2)(\a^2+k\b^2)}} \begin{pmatrix}
2\a^3+(k+1)\a\b^2\\
(k+1)\a^2\b+2k\b^3
\end{pmatrix}.
\end{equation}
Thus the first order derivatives  (together with the higher order ones) are  defined on the whole plane except the origin. 
The Hessian yields Finsler's metric of the form 
    \begin{eqnarray}\label{EE-metric}
    f_{ij}\!\!\!&\!\!\!\!\!\!=\!\!\!\!\!\!&\!\!\!\frac {1}{2\sqrt{(\a^2 + \b^2)^3(\a^2+k\b^2)^3}}\cdot\nonumber\\
   && \!\!\!\!\!\!\!\!\!\!\!\!
   \begin{pmatrix}
2\a^6+3(k+1)\a^4\b^2
+6k\a^2\b^4+(k^2+k)\b^6 & (k-1)^2\a^3\b^3\\
(k-1)^2\a^3\b^3 & (k+1)\a^6+6k\a^4\b^2+3(k^2+k)\a^2\b^4+2k^2\b^6
\end{pmatrix}.\nonumber\\
\end{eqnarray}
We calculate the determinant of this metric  by applying the Wolfram computation system. It is given by 
\begin{equation}\label{EE-det}
{\rm det}(f_{ij})= \frac {2(1+k)\a^4-(k^2-10k+1) \a^2\b^2 + 2k (k+1) \b^4}{{4(\a^2 + \b^2)(\a^2+k\b^2)}} 
\end{equation}
Observe that for $k=1$, we have here the standard  Euclidean metric $f_{ij}={\rm{diag}(1,1)}$, that is independent on the direction $v=(\a,\b)$.  
For $k=-1$, we have  the quartic  expressions $Q(v)=\a^4-\b^4$  with the corresponding Finsler metric  that we consider later. 
    \begin{figure}[H]\label{Fig2}

\centering
\includegraphics[width=0.38\textwidth]{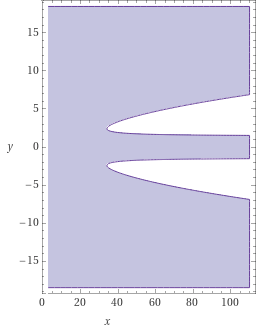}
\caption{In the plan with the axes $x=k$ and $y=\a/\b$, we depict the area (in blue) of positive determinant  ${\rm det}(f_{ij})>0$. In this case, the metric is Euclidean. Starting with the sufficiently big values of the parameter $k\sim 30$ the determinant is negative for small angles of the vector $v$. In this area, the space is Lorentzian.
}
\end{figure}

In Fig.(\ref{Fig2}), we present numerical presentation of the regions where the determinant (\ref{Eucl-det}) considering as a function of two variables  $k$ and $\a/\b$ is positive or negative. On the boundary curves, the metric is degenerated.  The asymptotic lines $y=\a/\b=\sqrt 2$ are visual.

For large values of the parameter $k$, the determinant approaches the asymptotic expansion  
\begin{equation}
    {\rm det}(f_{ij})\sim 
\left(\frac {2 \b^2- \a^2}{\a^2 + \b^2}\right) \, \frac k4\,.
\end{equation}
It means that for the large values of $k$ and sufficiently small tangent  of the vector $v$, such as $\b/\a<1/\sqrt 2$, there is  a region of the negative determinant, i.e., of the Lorentzian signature. In the exterior open domain, the metric is of the Euclidean signature. These domains are bounded by the strength lines where the metric is degenerated. 
    \begin{figure}[H]\label{Fig2}

\centering
\!\!\!\!\!\!\!\!\!\!\!\!\!\!\!\!\!\!\!\!\!\!\!\!\!\!\!\!\!\!\!\!\!\!\!\!\!\!\!\!\!\!\!\!\!\!\!\!\!\!\!\!\!\!\!\!\!\!\!\!\!\!\!\!\!\!\!\!\!\!\!\!\!\!\!\!\!\!\!\!
\begin{minipage}[t]{0.34\linewidth}
\includegraphics[width=0.85\textwidth]{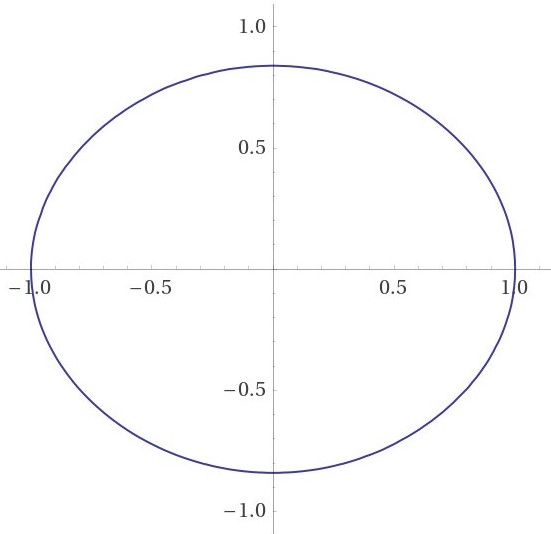}
\end{minipage}
\begin{minipage}[t]{0.34\linewidth}
\includegraphics[width=1.95\textwidth]{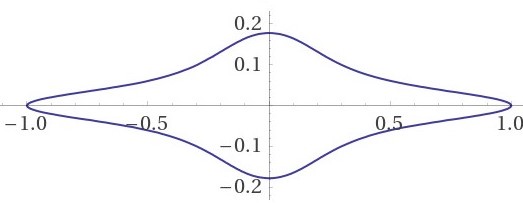}
\end{minipage}\\
\caption{The graphs present the solutions of the equation $Q(v)=1$ for different values of the parameter $k$. For small values of the parameter $k$, (the first graph) the curve is convex. For big values of the parameter $k$, (the second graph) the curve has  convex and concave parts.
}
\end{figure}

 Note a non-trivial behavior of Finsler's metric in the Lorentzian regime. Since 
 \begin{equation}
     f_{ij}v^iv^j=L(v)> 0,
 \end{equation}
the squared line element is positive for every $v\ne 0$. 
\subsection{Euclid $\times$ Lorentz}
Let us consider a quartic  expression
\begin{equation}\label{EL-quadric}
    Q(v)=Q(\a,\b)=(\a^2 + \b^2)(\a^2-k\b^2)\,,
\end{equation}
with a dimensionless {\it positive} numerical parameter $k$. 
Thus we have a product of a positive-definite quadratic and indefinite one.
The Lagrangian corresponding to this quartic 
\begin{equation}
    L(v)=L(\a,\b)=\sqrt{(\a^2 + \b^2)|\a^2-k\b^2|}\,,
\end{equation}
 The gradient of $L(v)$  takes the form
\begin{equation}
      L_{,i}=\frac {1}{\sqrt{(\a^2 + \b^2)|\a^2-k\b^2|}} \begin{pmatrix}
2\a^3+(1-k)\a\b^2\\
(1-k)\a^2\b-2k\b^3
\end{pmatrix}{\rm sgn}(\a^2-k\b^2).
\end{equation}
Thus the first order derivatives  (together with the higher order ones) are  defined on the whole plane except the pair of strength lines $\a^2=k\b^2$. 

The Hessian yields Finsler's metric of the form 
\begin{eqnarray}\label{}
    f_{ij}\!\!\!&\!\!\!\!\!\!=\!\!\!\!\!\!&\!\!\!\frac {1}{2\sqrt{(\a^2 + \b^2)|\a^2-k\b^2||^3}}\cdot\nonumber\\
   && \!\!\!\!\!\!\!\!\!\!\!\!
   \begin{pmatrix}
2\a^6+3(1-k)\a^4\b^2
-6k\a^2\b^4+(k^2-k)\b^6 & (k+1)^2\a^3\b^3\\
(k+1)^2\a^3\b^3 & (1-k)\a^6-6k\a^4\b^2+3(k^2-k)\a^2\b^4+2k^2\b^6
\end{pmatrix}.\nonumber\\
\end{eqnarray}
The determinant of this matrix takes the form 
\begin{equation}\label{EL-det}
{\rm det}(f_{ij})= \frac {2(1-k)\a^4-(k^2+10k+1) \a^2\b^2 - 2k (1-k) \b^4}{{4(\a^2 + \b^2)|\a^2-k\b^2|}} 
\end{equation}
\begin{figure}[H]\label{Fig2}
\centering
\includegraphics[width=0.55\textwidth]{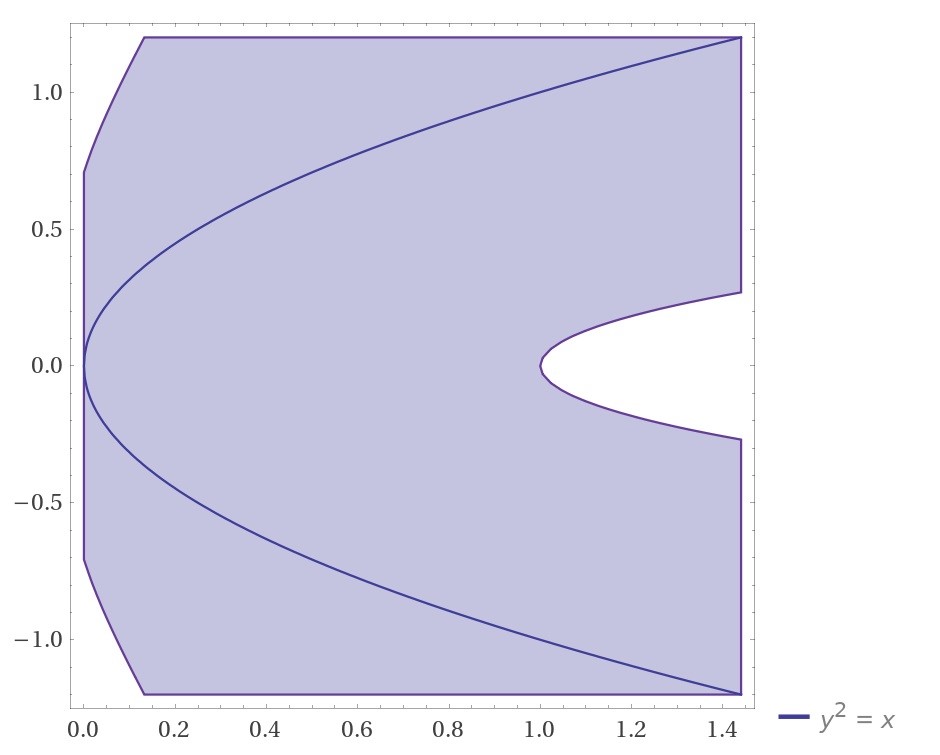}

\caption{
The solution of the inequality ${\rm det}(f_{ij})<0$ is presented in the blue area. The white area corresponds to the positive defined metric. The metric is  degenerated at the line $\a^2=k\b^2$.}
\end{figure} 

   \begin{figure}[H]\label{Fig2}

\centering
\qquad\begin{minipage}[t]{0.34\linewidth}
\includegraphics[width=1.55\textwidth]{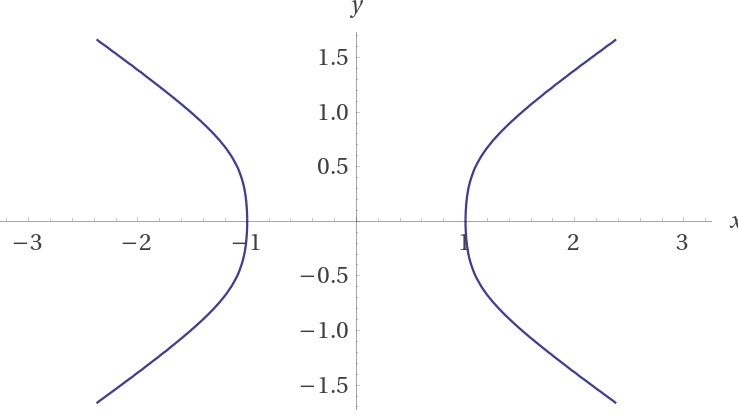}
\end{minipage}
\qquad\qquad\qquad\qquad\qquad\qquad\begin{minipage}[t]{0.24\linewidth}
\includegraphics[width=1.15\textwidth]{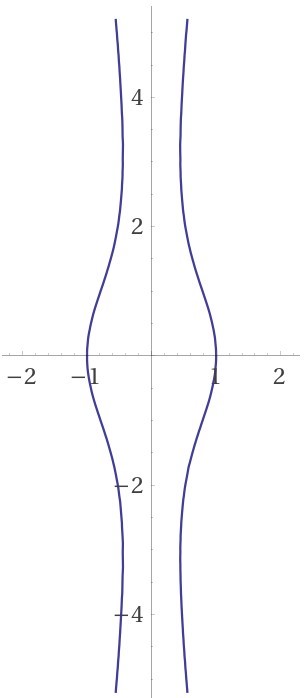}
\end{minipage}\\
\caption{Unit spheres in the  metric corresponding to the product quartic (\ref{EL-quadric}) for $k=2$ and $k=0.01$, respectively. The first graph is concave. On the second graph, the change of the convexity is visual.
}
\end{figure}

\subsection{Lorentz $\times$ Lorentz}
Let us consider a quartic  expression
\begin{equation}
    Q(v)=Q(\a,\b)=(\a^2 -\b^2)(\a^2-k\b^2)\,,
\end{equation}
with a dimensionless {\it positive} numerical parameter $k$. 
We have here a product of two indefinite factors.
Consider a Lagrangian for this quadric in the form
\begin{equation}
    L(v)=L(\a,\b)=\sqrt{|(\a^2 -\b^2)(\a^2-k\b^2)|}\,{\rm sgn}Q\,.
\end{equation}
The gradient of the Lagrangian reads
\begin{equation}
      L_{,i}=\frac {1}{\sqrt{|(\a^2 - \b^2)(\a^2-k\b^2)|}} \begin{pmatrix}
2\a^3-(1+k)\a\b^2\\
-(1+k)\a^2\b+2k\b^3
\end{pmatrix}{\rm sgn}\,Q.
\end{equation}
Using the general formula given in Eq.(\ref {Q-metric}) 
Finsler's metric is given by
\begin{eqnarray}\label{LL-metric}
    f_{ij}\!\!\!&\!\!\!\!\!\!=\!\!\!\!\!\!&\!\!\!\frac {{\rm sgn}\,Q}{2\sqrt{|(\a^2 - \b^2)(\a^2-k\b^2)|^3}}\cdot\nonumber\\
   && \!\!\!\!\!\!\!\!\!\!\!\!
   \begin{pmatrix}
2\a^6-3(1+k)\a^4\b^2
+6k\a^2\b^4-(k^2+k)\b^6 & (k-1)^2\a^3\b^3\\
(k-1)^2\a^3\b^3 & -(k+1)\a^6+6k\a^4\b^2-3(k^2+k)\a^2\b^4+2k^2\b^6
\end{pmatrix}.\nonumber\\
\end{eqnarray}


\begin{equation}\label{LL-det}
{\rm det}(f_{ij})= -\frac {2(k+1)\a^4+(k^2-10k+1) \a^2\b^2 + 2k (k+1) \b^4}{{4|(\a^2 - \b^2)(\a^2-k\b^2)|}} 
\end{equation}
  \begin{figure}[H]\label{Fig2}

\centering
\includegraphics[width=0.68\textwidth]{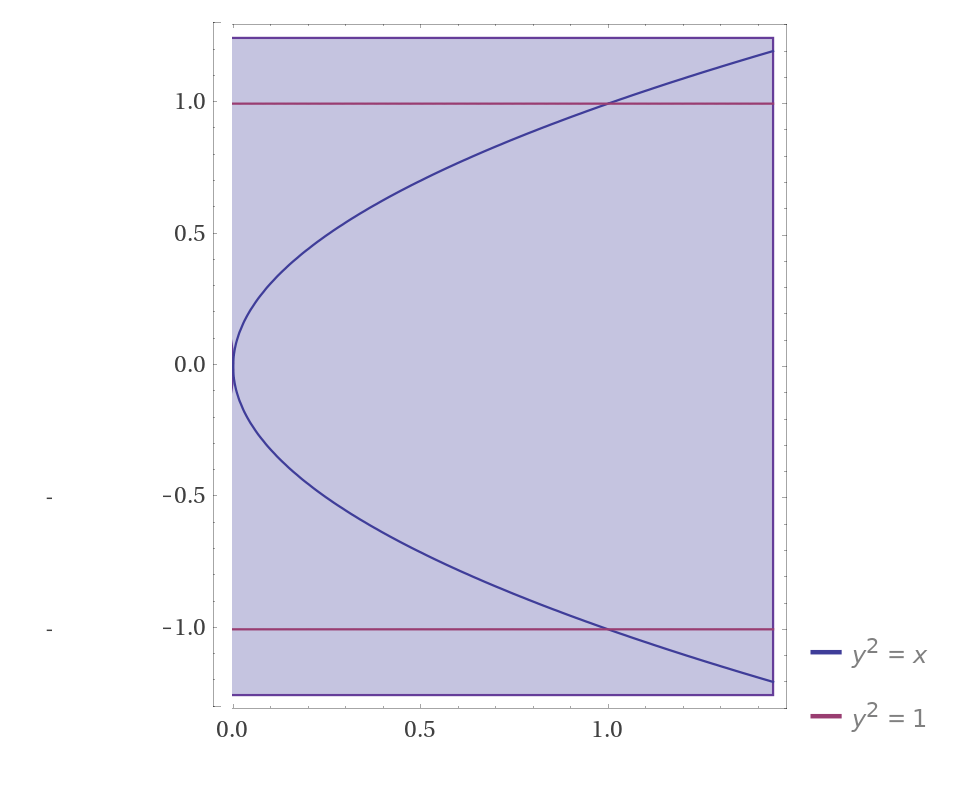}
\caption{In the plan with the axes $x=k$ and $y=\a/\b$, we depict the area (in blue) of the negative determinant  ${\rm det}(f_{ij})<0$. In this case, the metric is Lorentzian. The metric is degenerates on the lines $\a^2=\b^2$ and $\a^2=k\b^2$
}
\end{figure}
This determinant expression is strictly negative for all  positive values of the parameter $k$ and almost for all vectors $v$. The metric  degenerates on the hypersurfaces  $\a^2=\b^2$ and   $\a^2= k\b^2$. It means that the space is Lorentzian beyond the degenerate surfaces.  

  \begin{figure}[H]\label{Fig2}

\centering
\includegraphics[width=0.68\textwidth]{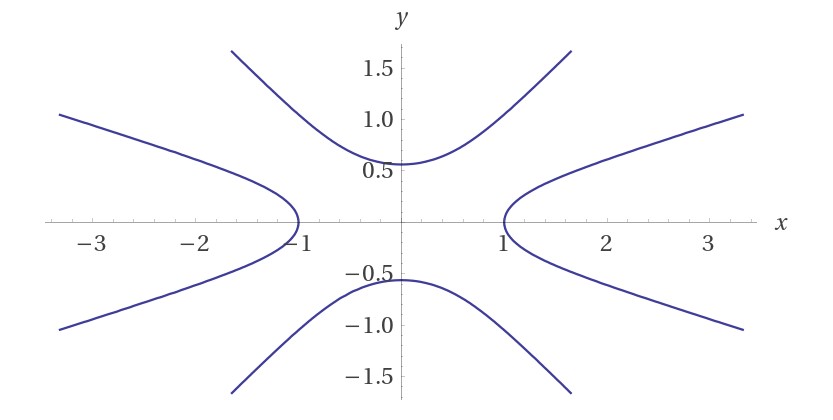}
\caption{For all values of the parameter $k$, the unit spheres are hyperbolas, as in the ordinary Lorentz space. 
}
\end{figure}

\end{document}